\documentclass[sigconf]{acmart}

\usepackage{ulem}
\usepackage{multirow}

\usepackage{booktabs} 
\usepackage{tikz}
\usepackage[theorems,skins]{tcolorbox}
\usepackage{etoolbox}
\usepackage{amsmath}
\usepackage{caption}
\usepackage{amsthm,amssymb}

\usepackage{empheq}
\usepackage{ulem}

\newcommand{\Policy}{{ \alpha}}

\newcommand{\eop}{\hfill{$\blacksquare$}}
\newtheorem{prop}{{\bf Proposition}} 
\newtheorem{thm}{Theorem}

\usepackage{bm}
\usepackage{framed}

\usepackage{ifthen}

\newcommand{\Detail}[1]{}
 
\newcommand{\lamS}{\lambda_{\sum}}
 \newcommand{\Rcc}{{\bm \Gamma}}
 
 \newcommand{\ignore}[1]{}
\newcommand{\IA}{\xi}

 \newcommand{\At}{{\mathcal A}}

\onecolumn

\acmConference[ ]{ }{ } {}

\begin{document}
 \title{Controlling Packet Drops to Improve Freshness of information}

  \author{$^1$Veeraruna Kavitha,  $^2$ Eitan Altman  and  $^1$ Indrajit Saha   \\  
 $^1$IEOR, IIT Bombay,  and  
 $^2$   INRIA France}

 \begin{abstract}

Many systems require frequent and regular updates of a certain information.  These updates have to be transferred regularly from the source to the destination. We consider scenarios in which an old packet becomes completely obsolete, in the presence of a new packet.  In this context,  if  a new packet arrives at the source while it is transferring a packet,
one needs to decide the packet to be dropped. New packet has recent information, but might require more time to transfer. Thus it is not clear as to which packet to be discarded, and this is the main focus of the paper.  Recently introduced performance metrics,  called average age of information (AAoI) and peak age of information (PAoI) of the information available at the destination,  are  the relevant performance measures.  These type of systems do not require storage  buffers, of size more than one, at the source queue.  We consider single source / multiple sources   regularly updating information to a single destination possibly over wireless channels to derive optimal drop policies that optimize the AAoI.  We showed that the state independent (static) policies like  dropping always the old packets or dropping always the new packets is optimal in many scenarios, among an appropriate set of stationary Markov policies.  
 We consider relevant games when multiple sources compete.  In many scenarios, the  non-cooperative solution `almost' minimizes the social objective, the sum of AAoIs  of  all the sources.

 \end{abstract}

  \maketitle

 \section{Introduction}
 
%
%
%
%

 Traditionally in queueing systems, the focus has been on delays and losses.  Recently, with the advent of applications demanding frequent and regular updates of 
  a certain information, the focus has been shifted towards the freshness of information. 
Timely updates of the information is an important aspect of  such systems, e.g,  sensor networks, autonomous flying vehicles etc. Many more such applications are mentioned in \cite{HowOften,M3,Aloha}. Most of the times the regular updates are transferred from the source of information to the destination using wireless communication systems.

 To measure the freshness of information, the concept of age of information (AoI), 
has been introduced as the duration between the observation 
time and the time when the latest received information sample  is generated \cite{HowOften}.
Peak age of information (PAoI) and Average age of Information (AAoI)  are the relevant performance measures, introduced 
recently in \cite{M3,HowOften}, and these differ significantly from conventional 
performance metrics, such as expected transmission delay, expected number of losses etc.

There has been considerable work in this direction since its recent introduction, we discuss a relevant few of them.
In \cite{HowOften} authors discuss the optimal rate of information generation that  minimizes the AAoI for various queuing systems. 
They showed  that  the  smallest  age  under  FCFS
can be achieved if a new packet is available exactly when the
packet in service finishes service.  
 In \cite{WirelessInfoFresh} the authors consider AoI only  for the packets waiting to be transferred/processed.  When the queue is empty their AoI is zero, their definition  accounts for the oldness of the information
waiting at the HoL (head of the line) of the link. 
In \cite{AoIMultiClass} authors study PAoI and   generalize the 
previously available results to the  systems with heterogeneous service time
distributions. 
The authors consider  
 update rates that  minimize the  maximum PAoI among all the sources.  
In \cite{Aloha} authors discuss   attempt probabilities for slotted aloha system that optimize  AAoI. We also consider Aloha, but our focus is on a radically different point, about dropping the `right' packets.  
%
%

Most of the work discussed above, considers lossless systems, where all the packets are transferred (possibly after some delays).  
 However often in systems which require regular updates of the same information, the old packet becomes obsolete once a new packet is available.  Thus it is more appropriate to consider lossy systems, where some packets are discarded,  while discussing the freshness of information.  
Lossy systems  have been a topic of interest   in  telecommunication networks, for many years now. 
For example, losses of information have become a tool to detect congestion (e.g., \cite{M1,M2}) in 
various transport control protocols (TCP). Lossy systems can model impatient customers/packets etc. 
Losses of packets may be due to buffer overflows  or  due to noise/interference in
wireless transmission. When the losses are due to buffer overflow,   one can 
decide the  packet to be dropped.  In this work we study the way in which the choice of
the packet to be dropped influences the freshness of the information.

If  a new packet arrives at source while it is in the middle of transferring a old packet, it appears upfront that the old packet has to be dropped. But 
if the transfer of the old packet is on the verge of getting completed, 
and if the new packets requires considerable time for transmission, 
it might be better to discard the new packet and continue the transmission of old packet.  
Further, the packet transfer times have large fluctuations when the packets are transferred through wireless medium. 
Thus it is not clear as to which packet is to be discarded. 

We showed that dropping the old packets  (always)  is optimal for  AAoI, 
when the packet transfer times are distributed   according to    exponential   or     hyper exponential distribution.  
This is a static  policy as the drop decision does not depend upon the state of the system, but   is optimal among all the   stationary Markov and randomized  (SMR) policies. The SMR dynamic policies   depend upon the age of information at an appropriate decision  epoch. 
We also establish   certain conditions under which  dropping the new packets is optimal among SMR policies.
For  transfer time distributions like uniform,   Weibull, Poisson, log-normal etc.,   dropping old/new packets  is optimal based on the parameters. 
With the aid of numerical computations we showed  for almost all cases that, either of the two static policies are almost optimal. 

%

The second part of this work considers  multiple sources transferring regular updates of their  information to a common destination. 
An important conclusion of this study among others  is: the natural dropping of packets (only ones with good channels are selected)  in CSMA  (Carrier-sense multiple access) environment is optimal for a social choice function. We observe that the profile of  attempt probabilities that form the Nash equilibrium  for Random CSMA environment also  minimizes the sum of the AAoI of all the agents.   Thus the agents have no incentive to deviate from the cooperate solution.


\ignore{ 
Lossy systems  and congestion control have been a topic of interest   in  telecommunication networks, for many years now. 
Losses of packets may be due to buffer overflows  or to a noise or
interference in
wireless transmission. When losses are due to buffer overflow then one can 
decide which packet should be dropped on an overflow event. This is the
subject of this work. In this work we study the way in which the choice of
the packet to be dropped influences the freshness of the information. }
\ignore{
In \cite{M1,M2} the authors have studied the impact of controlled 
losses on the system throughput when TCP connections compete in grabbing
bandwidth at a bottleneck link. A fluid model is used there to model
the throughput of TCP. Optimal dropping policies are derived.
}

In section \ref{sec_syswithlosses} we introduce the subject and the two static policies, dropping always the new packets and dropping always the old packets. 
Section \ref{sec_control}   considers optimal drop policies for single source. Sections \ref{sec_multiple} (with zero storage) and \ref{sec_multiple_storage} (with one storage)  consider  multiple sources.

 
 {\bf Notation}  We have cycle numbers,  sub-cycle numbers,  counters for events within a cycle/sub-cycle. We also have indices for sources. When it is not important, we suppress  some of these numbers to simplify the notation. For example, when a quantity across different cycles is identically distributed and when one considers its expectation we do not mention the cycle number. 
 
 
 \section{System with losses, fresh updates} 
 \label{sec_syswithlosses}

Consider  source(s) sending regular updates of a certain information to  destination(s). The information update packets arrive at any source according to a Poisson process with rate $\lambda$.  The packets are of the same size but the transfer times might vary based on the medium.  A  source  requires IID (independent and identically distributed) times   $\{ T_i\}$   to deliver the packets to the destination, which are equivalently the job times in the queue.  
Our focus is on measures related to the freshness of information available at the destination for different systems.  
We begin with single source-destination pair and consider multiple sources sharing wireless resources, using CSMA protocol, in later sections. 

 \subsubsection*{Age/Freshness of information} 
The age of information (AoI), from the given source and at the given destination,  at time $t$ is defined 
$$
G(t) := t -  r_t,
$$where $r_t$ is the time at which the last successfully received  packet (at destination) before time t, is generated.  Each packet is generated at a fixed time $\delta$ before its arrival instance to the source queue and one can neglect $\delta$, as it is mostly a constant value.     Our aim is to study the (time) average age of information (AAoI), defined as below\footnote{Limit exists almost surely in all our scenarios.}:
\begin{eqnarray}
\label{Eqn_AAoI}
{\bar a} :=  \lim_{T \to \infty}  \frac{\int_0^T G(t) dt }{T}.
\end{eqnarray}

As already mentioned,  we consider freshness of information in a  lossy system,  and our focus is on the  packet to be dropped when there are two simultaneous packets.
 We initially consider  systems with zero storage.  We begin with analysis of the system that drops new packets, when busy.

\subsection{Drop the new packets (DNP)} 
\begin{figure}[h]
\begin{center}

\vspace{-6mm}
\includegraphics[scale=.5]{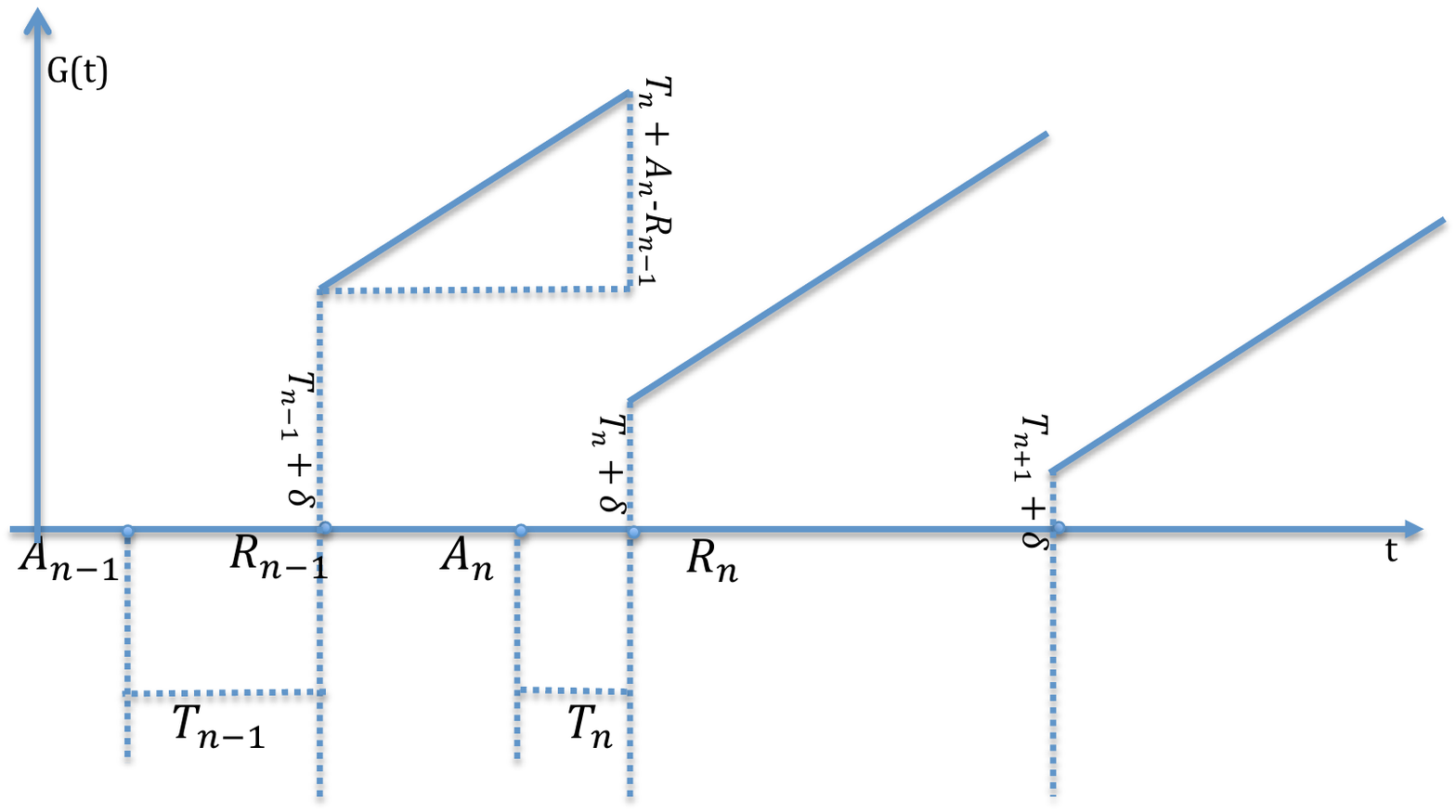}
\vspace{-105mm}
\caption{DNP scheme,  Renewal cycle \label{fig_DNP}}
\end{center}
\vspace{-3mm}
\end{figure}

The source does not stop/interrupt transmission of any packet. If a new update packet arrives, in between transmission, it is dropped. Once the transfer is complete (after random time $T$), the source waits for new packet, and starts transmission of the new packet immediately after. And this continues (see Figure \ref{fig_DNP}).  

  The age of the information $G(t)$  grows linearly with time at unit rate, at all time instances, except for the one at which a packet is just received at the destination. 
   At that time epoch the age drops to 
$
T_k  ,
$ because:  a)  $T_k$ is the time taken to transfer the (new) packet from source to destination, after its arrival at the source queue;  and  b) this represents the age of the new packet at destination. 

Thus we  have a renewal process as in   Figure \ref{fig_DNP}.  Here $\{R_k\}$ are the epochs at which a message is transferred successfully and would become the renewal instances, while $\{A_n\}$ are the arrival instances of the packets (at the source and of those transferred) governed by Poisson point process (PPP).  Let $\{\xi_k\}_n$ represent the corresponding inter-arrival times.
As seen from the figure, the age of the information is given by sawtooth kind of waveform. Further, clearly, the alternate renewal cycles are independent of one another. 
Thus one can apply renewal reward theorem\footnote{One can apply to even cycles and odd cycles separately and consider the average of the two.}  (RRT) and the long run time average of the age of the information defined in (\ref{Eqn_AAoI}) almost surely (a.s.) equals:
\begin{eqnarray}
{\bar a}& =&     \frac{E \left [ \int_{R_{k-1}}^{R_k} G (s) ds \right ]  }{ E[ R_k - R_{k-1}]}, \mbox{ and with } G_k := G(R_{k}).
 \nonumber
  \\
&
=&  \frac{E[G_{k-1} ( R_k - R_{k-1}) ]  + 0.5  E[(R_k - R_{k-1} )^2] }{ E[R_k - R_{k-1}] } \mbox{ a.s.  } \nonumber  \\
&=&E[G_{k-1} ]  + 0.5  \frac{ E[(R_k - R_{k-1} )^2] }{ E[R_k - R_{k-1}] } \mbox{ a.s.  } 
  \hspace{2mm} 
\label{Eqn_AAoI_by_RT_limit}
\end{eqnarray}
The last line follows by independence (see Figure \ref{fig_DNP})  and   memoryless property of PPP.
For DNP scheme, $G_{k-1} =  T_{k-1}  $,  and  we have:
$$
{\bar a}_{DNP} 
 =  E[T_{k-1}]   + \frac{1}{2} \frac{ E[( T_k + \IA_k )^2] }{ E[ T_k + \IA_k] } \mbox{ a.s.,}
$$where $\IA_k$,  the inter-arrival time, is exponentially distributed with parameter $\lambda$ and is independent of the transfer times $T_k, T_{k-1}.$
Simplifying 
\begin{eqnarray}
\label{Eqn_age_DNP}
{\bar a}_{DNP} 
& =&  E[T] + \frac{1}{2 (1/\lambda + E[T] )}   \left ( \frac{2}{\lambda^2} + \frac{2 E[T] }{\lambda}  + E[T^2] \right ) \nonumber \\
 &=&   E[T] +  \frac{1}{\lambda}  +  \frac{E[T^2]}{2 E[T]}   \frac{\rho}{1+ \rho}  \mbox{ with } \rho := \lambda E[T].
\end{eqnarray}
{\it We are considering identical quantities, hence we drop the time index $k$ when it is not important.}

\subsection{Drop the old packets  (DOP)}  

When source receives a new message, the ongoing transfer (if any) of the old packet is stopped and the old packet is dropped. The source immediately starts transfer of the new packet.   The new message would imply a more fresh information, but might also imply longer time (because we now require the transfer of the entire message)  before the    information at the destination is updated.   However the variability in transfer times $\{T_k\}$ might imply interruption is better for average freshness under certain conditions, and we are keen in studying this  aspect. 

 The renewal points will again be the instances at which a message is successfully received. But  note that only when a message transfer is not interrupted by a new arrival, we have a  successful   message   reception.  Thus the renewal cycles in Figure (\ref{fig_DNP}) get prolonged appropriately.  Let $A_{k,1}$  be the first arrival instance after the $(k-1)$-th renewal epoch $R_{k-1}$. Let  $\IA_{k, 0}$  be the corresponding inter-arrival time (which is exponentially distributed). 
Its service (i.e., message transfer) starts immediately and let $T_{k,0}$ be the job size, or the  (random) time required to transfer this message . In case a second arrival occurs (after inter-arrival time $\IA_{k,1}$) within this service, we start the service of the new packet by discarding the old one. This happens with probability  $1-\gamma$ where 
\underline{$\gamma := P(T_{k,0} \le  \IA_{k, 1}) $}.  The renewal cycle is completed after second transfer,  in case the second message transfer is not interrupted.  The second can also get interrupted,  independent  of previous interruptions and  once again with the same probability $1-\gamma$,  because of IID nature of  the transfer times and the inter arrival times.  If second is also interrupted the transfer of the third one starts immediately and this continues till a job is not interrupted (i.e., with probability $\gamma$). And then the renewal cycle is completed.

 Once again the alternate cycles are IID, RRT can be applied  to AAoI given by (\ref{Eqn_AAoI})  and AAoI is given by equation (\ref{Eqn_AAoI_by_RT_limit}).   However the renewal cycles $\{R_k - R_{k-1}\}_k$ are  more complex now, and we proceed with deriving their moments.     
  The $k$-th renewal cycle can be written precisely as below,  using the arrival sequence $\{ \xi_{k, i}\}_{i\ge 0}$ and transfer times sequence $\{ T_{k, i}\}_{i\ge 0}$  belonging to $k$-th renewal cycle: \vspace{-4mm}
\begin{eqnarray}
  {R_c}_k \hspace{-2mm}&\hspace{-1mm}:=\hspace{-1mm}& \hspace{-2mm} R_{k} - R_{k-1} =  \xi_{k,0} +  \hspace{-2mm}\sum_{i=1}^{{\mathcal N}_k -1} \xi_{k, i}  + T_{k, {\mathcal N}_k-1}  \label{Eqn_Rc_a.s.} \\
  &=& \nonumber  
 \xi_{k,0} + \Rcc_k, \ \mbox{ with} \hspace{3mm} 
\\
   \Rcc_k \hspace{-2mm}&\hspace{-1mm}:=\hspace{-1mm}& \hspace{-3.5mm} \sum_{i=1}^{{\mathcal N}_k -1}  \xi_{k, i}  + T_{k, {\mathcal N}_k-1}, \mbox{ and } \label{Eqn_Rcc}   \\
   {\mathcal N}_k  \hspace{-2mm}&\hspace{-1mm}:=\hspace{-1mm}&   \inf  \big  \{  i \ge 1:  \xi_{k, i}  > T_{k, i-1}  \big \}.  \hspace{3.8mm } \nonumber \end{eqnarray}
In the above ${\mathcal N}$ is the number of interruptions before successful transfer, and it  is geometrically distributed with parameter $1-\gamma$ 
and $\Rcc$ (given by (\ref{Eqn_Rcc})) is the time to taken complete one packet transfer, in the midst of interruptions by  new arrivals.  The above random variables are specific to a given renewal cycle, but are also IID across different cycles.    Further,  $G_k = G(R_k)$ is now  a  `special'   transfer time  (represented by $\underline {T}$):  one which is  not interrupted.  Thus
\begin{eqnarray}
\label{Eqn_Gk_DOP}
G_k =  {\underline T}_k :=  T_{k, {\mathcal N}_k-1} \mbox{,  }  E[G_k] =E[ {\underline T}_k] =  E[T | T \le \xi] \mbox{ for any } k. \end{eqnarray}
Hence
 the AAoI of DOP scheme (again by independence of alternate cycles) equals (see (\ref{Eqn_AAoI_by_RT_limit})):
\begin{eqnarray}
\label{Eqn_AAoI_DOP} 
{\bar a}_{DOP}&=& 
E[T | T \le \xi]     + \frac{   E[(\xi + \Rcc )^2] }{2  E[\xi + \Rcc] }
 \mbox{ almost surely (a.s.). } 
\end{eqnarray} 
Thus to complete the analysis we require  the first two moments of   $\Rcc$ (see equation (\ref{Eqn_Rc_a.s.})).   
 We compute  $E[\Rcc]$ and $E[\Rcc^2]$, by  conditioning on $\xi_{k,1}$, $T_{k, 0}$ as below:
   \begin{eqnarray*}
  E[\Rcc_k] &=&  E \left [ \Rcc_k   \  \bm{;}  \  \xi_{k,1} > T_{k, 0 } \right ]      +
    E  \left [  \Rcc_k  \ \bm{;} \   \xi_{k,1} \le T_{k, 0} \right ]    \\
  &=&  E \left [ T_{k, 0}  \  \bm{;}  \  \xi_{k,1} > T_{k, 0 } \right ]    
    +  E  \left [ \xi_{k, 1} + \tilde {\Rcc}  \  \bm{;}  \  \xi_{k,1} \le T_{k, 0} \right ]    \\
  && \hspace{-14mm} = E \left [ T_{k, 0}    ;    \xi_{k,1} > T_{k, 0 }  +   \xi_{k, 1}  \  \bm{;}  \ \xi_{k,1} \le T_{k, 0} \right ]   
    +  E\left [  \tilde {\Rcc}    \  \bm{;}  \ \xi_{k,1} \le T_{k, 0} \right ]   ,
  \end{eqnarray*}where $\tilde \Rcc$ is an IID copy of $\Rcc_k$, which is independent of $T_{k, 0}$ and $\xi_{k, 1}$. 
After simplifying and further  conditioning on $T$:
 \begin{eqnarray}
   E(\Rcc)  &=&\frac{  E \left [ T     ;    \xi  > T  +   \xi   \  \bm{;}  \ \xi \le T \right ]   } { P( T  \le  \xi  ) } \nonumber \\
   &=&  \frac{  E[Te^{-\lambda T} ] +  ( 1- E[e^{-\lambda T} ]  ) /  \lambda - E[T e^{-\lambda T} ]  }{ P( T  \le  \xi  )  } \nonumber \\
   &=& \frac{ 1- \gamma  }{\lambda \gamma  }. 
\label{Eqn_ERC} 
\end{eqnarray}
Using exactly similar logic:
\begin{eqnarray*}
E[\Rcc^2] = E[\min \lbrace T_0 , \xi _1 \rbrace^2] + E[{\tilde \Rcc}^2]  (1-\gamma)   + 2E[{\tilde \Rcc} ]E[  \xi _{k,1}   \  \bm{;}  \  T_{k, 0} >  \xi_{k,1}).
\end{eqnarray*}
 Using (\ref{Eqn_ERC})
\begin{eqnarray}
E[\Rcc^2] &=&\frac{E[min \lbrace T_0 , \xi _1 \rbrace^2]  + 2E[\Rcc]E[ \xi _{k, 1}    \  \bm{;}  \ T_{k, 0} >  \xi_{k, 1} \rbrace] }{\gamma } \nonumber \\
%
&& \hspace{-18mm} =  \frac{ 2 (1-  \gamma )  }{\lambda^2 \gamma } - \frac{2 E[T e^{-\lambda T}]}{ \lambda\gamma }    \nonumber  
  +\frac{ 2 E[\Rcc]  }{\gamma } \left (\frac{1-\gamma }{\lambda} -  E[Te^{-\lambda T} ] \right )  \nonumber \\
%
&& \hspace{-18mm} =  \frac{ 2 (1- \gamma )  }{\lambda^2 \gamma } - \frac{2 E[T e^{-\lambda T}]}{ \lambda \gamma^2 }    +
\frac{2 (1-\gamma)^2}{\lambda^2 \gamma^2}  
= \  \frac{ 2 (1- \gamma )  }{\lambda^2 \gamma^2 } - \frac{2 E[T e^{-\lambda T}]}{ \lambda \gamma^2 }  .    \hspace{2mm}
\label{Eqn_ERC_sqr}
\end{eqnarray}
 Using   (\ref{Eqn_ERC}) and (\ref{Eqn_ERC_sqr}), the first two moments of the renewal cycle are:
 \begin{eqnarray}
 \label{Eqn_ERC_two_moments}
 E[R_c]  &=& \frac{1}{\lambda}  + \frac{1-\gamma}{\lambda \gamma} = \frac{1}{\lambda \gamma}  \mbox{ and }  \\
 E[R_c^2 ]  &=&  E \left [\xi_{k, 0}^2  + 2 \xi_{k, 0} \Rcc_k + \Rcc_k^2 \right  ] \nonumber  
 \Detail{\\ &=&  \frac{2}{\lambda^2} + \frac{2 (1-\gamma) }{\lambda^2 \gamma} +  \frac{ 2 (1- \gamma )  }{\lambda^2 \gamma^2 } - \frac{2 E[T e^{-\lambda T}]}{ \lambda \gamma^2 }      . \nonumber \\ 
 &=& }{ \  =  \ }  \frac{2}{\lambda^2 \gamma^2}   - \frac{2 E[T e^{-\lambda T}]}{ \lambda  \gamma^2 }.
 \end{eqnarray}
Substituting the above into (\ref{Eqn_AAoI_DOP})  the AAoI for DOP scheme equals:
\begin{eqnarray}
\label{Eqn_DOP_AAoI}
{\bar a}_{DOP}  =  \frac { E[T e^{-\lambda T} ]}{\gamma }  + \frac{1}{\lambda \gamma}   - \frac{ E[T e^{-\lambda T}]}{   \gamma }  
=  \frac{1}{\lambda \gamma}.
\end{eqnarray} 
Thus the AAoI with DOP scheme equals the expected renewal cycle, while that with DNP scheme is strictly bigger than the expected renewal cycle (see (\ref{Eqn_age_DNP})).   It is not guaranteed that the expected renewal cycle with DOP scheme is smaller than that with DNP scheme. Thus it is not clear upfront as to which scheme is better.   But it is equally (or more)  important to understand if any scheme with controlled drops can perform better than the two schemes.

\section{Controlled drops: Single source}
\label{sec_control}

In the previous  section  two `extreme' and static schemes are considered: in one all the old packets are dropped while in the other all the new packets are dropped. In this  section we investigate if there exist a better  scheme with partial/controlled drops.  We also study the conditions under which DOP is better than DNP.

With message successful transfer epochs $\{R_k\}_k$ as the decision epochs, we consider a {\it dynamic decision about the (DOP/DNP) scheme to be used}. The dynamic decision depends upon  
the state, the age of 
information $G_k$,  at the decision epoch $R_k$.  
We initially restrict ourselves to  special type of {\it dynamic  policies, called threshold policies:}  DNP scheme is selected if age ($G_k$) is above a threshold (say $\theta\ge 0$) and DOP is selected other wise.  

\Detail{
 Let $\{ R_k\}_k$ again represent the renewal epochs\footnote{Note that these cycles are no more independent, nor are they identically distributed, however we continue to refer them as renewal cycles.}, i.e., the epochs of successful message transfers, $\{ {R_c}_k \}_k$ represent the length of corresponding renewal cycles and  $G_k = G(R_k)$ represent the age of information at destination at the renewal epoch $R_k$, as before.  
 }{}
 To be precise any
  $\theta \ge 0$  represents a controlled scheme,  and 
 DOP  scheme is chosen for (entire)  renewal cycle  starting at  $R_k$,  if   $G_k  < \theta$.  Otherwise,  DNP scheme\footnote{The analysis would be similar and the results are the same if the schemes are reversed, i.e., if DOP scheme is chosen with $G_k > \theta$. We consider much more generalized policies in sub section \ref{sec_SMR}} is chosen. 
 %
With DNP scheme,  new packets are dropped (other than the first one in that renewal cycle) till the message transfer is complete. 
 With DOP decision, old packets are dropped and transmission of new packet starts immediately,  whenever the former  is interrupted. This continues till   a message is transferred completely. 
  Further dropping of  (old/new) packets depends upon the decision at  the next decision epoch. 
 
 \subsection{Transitions}
   In contrast to the previous subsections, the  length of renewal cycles $\{{R_c}_k\}_k$ are no more identically distributed. The distribution of ${R_c}_k$ depends upon the scheme chosen  at the decision epoch $R_k$.  
    It is easy to observe that the length of the renewal cycle ${R_c}_k$ does not depend upon the absolute value of  state $G_k$, but only upon the state dependent binary (DOP/DNP) decision. Thus the distribution of  ${R_c}_k$  can be one among two types and is precisely given by (see  (\ref{Eqn_Rc_a.s.})):
\begin{equation}
\label{Eqn_transitions_Rc}
  {R_c}_{k+1}= \left \{
  \begin{aligned}
    & \xi_{k+1, 0} +T_{k+1, 0}, &&   \mbox{with  DNP } (G_k > \theta)  \\
    & \xi_{k+1, 0} +  \Rcc_{k+1}   &&   \text{else.}
  \end{aligned} \right.
\end{equation} 
The state update has similar transitions\Detail{,  i.e., the transition do not depend directly upon the previous state but only via  the action chosen as below}{} (see (\ref{Eqn_Gk_DOP})): 
 \begin{equation}
 \label{Eqn_transitions_G}
  G_{k+1}= \left \{
  \begin{aligned}
    & T_{k+1, 0}, &&  \mbox{ with DNP  } (G_k  > \theta )  \\
    & {\underline T} _{k+1}  , && \text{ other wise. }
  \end{aligned} \right.
\end{equation} 
Observe that $\theta  = 0$ implies DNP,   while DOP is obtained by considering $\theta \to \infty.$   {\it For ease of notation  we say $\theta = \infty$  when DOP is selected for all  $G_k$.}  

 \subsection{Analysis}
For every $\theta$,  the random variables $\lbrace G_k\rbrace_k$   and $\lbrace {R_c}_k \rbrace_k$ constitute a  Markov chain  and with the help of this Markov chain we will compute the AAoI.   One can rewrite AAoI (\ref{Eqn_AAoI}) as:
$$
{\bar a} (\theta)  = \lim_{k \to \infty}  \frac{ \int_0^{R_k}  G(t) dt }{R_k } \mbox{ a.s., }
$$because   $R_k \to \infty$ a.s. as $k \to \infty$, and this is because 
$$R_k  = \sum_{l\le k} {R_c}_l \ge \sum_{l\le k } \xi_{l, 0} \mbox{ for all } k \mbox{ and }  \sum_{l\le k } \xi_{l, 0} \stackrel {k \to \infty}{\to} \infty \mbox{  a.s. } $$
Thus  \vspace{-4mm}
\begin{eqnarray}
\label{Eqn_SLLN_reqd}
{\bar a}  (\theta) \hspace{-2mm}& \hspace{-1mm}= \hspace{-1mm}&\hspace{-2mm} \lim_{k \to \infty}   \frac{ \sum_{l \le k} \int_{R_{l-1}}^{R_l}  G(t) dt }{R_k }   \nonumber \\ 
 \hspace{-2mm}& \hspace{-1mm}= \hspace{-1mm}&\hspace{-2mm}   \nonumber
  \lim_{k \to \infty} 
  \frac{ \sum_{l \le k} \int_{R_{l-1}}^{R_l}  G(t) dt }{k} \frac{k }{\sum_{l \le } {R_c}_l } ,  \nonumber   \\ 
 \hspace{-2mm}& \hspace{-1mm}= \hspace{-1mm}&\hspace{-2mm} \lim_{k \to \infty} 
  \frac{ \sum_{l \le k} \left ( G_{l-1} {R_c}_l + 0.5 {R_c}_l^2 \right )  }{k} \frac{k }{\sum_{l \le } {R_c}_l } . \hspace{3mm} \label{Eqn_SLLN_reqd2}
\end{eqnarray}

 As already discussed, the distribution of ${R_c}_k$ (for any $k$)  can be of two types depending only   upon the event $\{ G_k < \theta\}$ (see (\ref{Eqn_transitions_Rc}). In exactly a similar way the stationary distribution  ${{R_c}_*}$ depends only upon the stationary event $\{G_* < \theta\}$. Thus it suffices to obtain the stationary distribution of 
 $\{G_k\}_k$.  In fact the transitions of $\{G_k\}$ given by (\ref{Eqn_transitions_G}) also depend only upon the events  $\{G_{k-1} < \theta\}$. Thus it further suffices to study the two state Markov chain $X_k := 1_{ \{ G_k <  \theta \} }$ ($1_A$ is the indicator of the event $A$) and the rest of the random quantities can be studied using this two state chain.  The Markov chain has the following evolution 
\begin{equation}
  X_{k+1}= \left \{
  \begin{aligned}
    & 1_ {\lbrace T_{k+1, 0} <\theta \rbrace}                    && \text{if}\ X_k=0,  \\
    & 1_{ \lbrace {\underline T} _{k+1}  < \theta \rbrace}  && \text{else.} 
  \end{aligned} \right.
\end{equation} 
When $ \theta =  \infty  $, $X_k \equiv 1$ for all $k$.
 The  transition probabilities  (with $\theta \ne \infty$) are:
 \begin{eqnarray}
 \label{Eqn_transition_two}
 P(X_{k+1} = x' | X_{k}  = x)  
 & =& \left \{
 \begin{array}{llll}
 p_\theta                                                                       & \mbox{ if }  x = 0, \  x' = 1  \nonumber  \\
q_\theta  & \mbox{ if }  x = 1,  \  x' = 0  \nonumber  \ \  \mbox{ where }
 \end{array}   \right .  \nonumber  \\
   p_\theta &:=& P(T< \theta)  \mbox{ and }     \  \\
 q_\theta &:=& P({\underline T} > \theta)  = P (T > \theta \big | T \le  \xi) . \nonumber  \hspace{3mm}
\end{eqnarray}
This chain has unique stationary distribution   given by: 
\begin{eqnarray}
\label{Eqn_pi}
\pi_\theta (0) = \frac{q_\theta }{q_\theta +p_\theta } 1_{\theta \ne  \infty} = 1-\pi_\theta (1), \mbox{ and } P(X_* = 0) = \pi_\theta(0). \end{eqnarray}
Let $X_*$, $G_*$ and ${R_c}_*$ represent the random quantities corresponding to  stationary distributions of $X_k$, $G_k$ and ${R_c}_k$ respectively.
The stationary distribution of the remaining quantities is  dictated by that of $\{X_k\}$:  for example the stationary distribution of  $G_*$ is the same as that of $T$, a typical transfer time when $X_* = 0$ and equals that of ${\underline T}  = T | T \le \xi$ (the conditional distribution)  when $X_* = 1.$

The Markov chain $\{X_k\}$ is clearly ergodic,  the rest of the stationary random quantities  ${R_c}_*$, $G_*$ depend just upon $X_*$, hence  strong law of large numbers (SLLN) (e.g., \cite{Meyn}) can be applied\footnote{
One can not apply the usual renewal theory based analysis, as  the process is (the odd/even cycles are also)  Markovian and \underline{can not be modelled as a Renewal process}, with IID renewal cycles. 
}  separately to the numerator and denominator of (\ref{Eqn_SLLN_reqd})   to obtain:
$$
{\bar a}(\theta) =  \frac{E_{\pi_\theta} [ G_{l-1} {R_c}_l ]  + 0.5 E_{\pi_\theta} [  {R_c}_l^2  ] \ }{  E_{\pi_\theta} [  {R_c}_l  ]   } \mbox{ a.s. },
$$ 
where  $E_{\pi_\theta}  [ \cdot ]$ is  the stationary expectation.
It is easy to verify  (see (\ref{Eqn_transitions_Rc})) by appropriate conditioning that:

\vspace{-4mm}
{\small \begin{eqnarray*}
E_{\pi_\theta} [ G_{l-1} {R_c}_l ] \hspace{-2mm} &\hspace{-1mm}=\hspace{-1mm} &  \hspace{-2mm}  E_{\pi_\theta} [ G_{l-1} {R_c}_l ; X_{l-1}  = 1] 
 +  E_{\pi_\theta} [ G_{l-1} {R_c}_l ; X_{l-1} = 0]   \nonumber 
\\
&& \hspace{-13mm} = \   
\left (\frac{1}{\lambda}  + E[T] \right ) E[G_*  ;   G_* > \theta ]   +  E[G_* \  ; \  G_* < \theta ] \left (\frac{1}{\lambda}  + E[\Rcc] \right ) \mbox{{\normalsize and} }   \\
 E[G_*  ;   G_* > \theta ] \hspace{-2mm} &\hspace{-1mm}=\hspace{-1mm} &  \hspace{-2mm}    E[T ; T > \theta ] \pi_\theta (0) + E[ T ; T > \theta  | T\le \xi ]  \pi_\theta (1)  .
\end{eqnarray*}}
Using similar logic, 
\begin{eqnarray}
 E_{\pi_\theta} [ G_{l-1} {R_c}_l ]  + 0.5 E_{\pi_\theta} [  {R_c}_l^2  ]   \hspace{-28mm}  \nonumber \\
     &= & d_n E[G_*  ;   G_* > \theta ]     \nonumber
      + d_o  E[G_* \  ; \  G_* < \theta ]      + 0.5   E[{R_c}_*^2  ]   \nonumber \\    
      \Detail{
  && \hspace{-34mm} = \   \left (\frac{1}{\lambda}  + E[T] \right ) \bigg  (  E[T ; T > \theta ] \pi_\theta (0) + E[ T ; T > \theta  | T\le \xi ]  \pi_\theta (1)    \bigg  )  \nonumber   \\
  && \hspace{-28mm}+  \left (\frac{1}{\lambda}  + E[\Rcc] \right )    \bigg  (  E[T ; T \le  \theta ] \pi_\theta (0) + E[ T ; T \le  \theta  | T\le \xi ]  \pi_\theta (1)    \bigg  )   \nonumber  \\
  &&\hspace{-28mm}+  0.5   E[(T + \xi)^2]  \pi_\theta  (0) + 0.5 E[ (\xi + \Rcc)^2 ] \  \pi_\theta (1),   \nonumber \\}{}
   & = & \beta_\theta (0) \pi_\theta (0) + \beta_\theta (1) \pi_\theta (1)  ,   \label{Eqn_Num_limit}  \\
 E_{\pi_\theta} [  {R_c}_l  ]  &=&     d_n \pi_\theta (0) +  d_o    \pi_\theta (1), \nonumber
   \end{eqnarray}
with the following definitions:
  \begin{eqnarray*}
 \beta_\theta (0)  &:=& d_n   E[T ; T > \theta  ] +   d_o     E[T ; T \le  \theta ] 
  +  0.5   c_n, \\
  \beta_\theta (1)  &:=&   d_n  E[T ; T > \theta  | T\le \xi  ] 
+   d_o    E[T ; T \le  \theta  | T\le \xi  ] +  0.5   c_o  , \\
            c_n &:= & E[(\xi +T)^2] , \hspace{10mm}  \ c_o := E \left [\left (\xi + \Rcc \right )^2 \right ]  \mbox{ and }  \\
 d_n &:=& E[T + \xi]  , \hspace{14mm} \   d_o := E [\xi +  \Rcc ]   .
\end{eqnarray*}
 
 Thus the AAoI equals
 \begin{equation}  \label{Eqn_AAoI_theta}
 \bar{a}(\theta) = \frac{\beta_\theta (0)  \pi_\theta(0) + \beta_\theta (1)  \pi_\theta(1)}{ d_n \pi_\theta (0) +  d_o \pi_\theta(1)}
   {\mbox{   a.s.}}
 \end{equation}
 Here AAoI is defined using the stationary expectation  ($\pi_\theta$), while  (\ref{Eqn_AAoI_by_RT_limit}) is for the case with (alternate) IID cycles.

 \subsection{Optimal $\theta$}
 We are interested in  optimal threshold, $\theta^*$ and  hence consider:
 \begin{eqnarray}
\label{Eqn_obj}
   \min_{ \theta \ge 0 }  \bar{a}(\theta).
   \end{eqnarray}  
   The objective function depends upon $\theta$ in a complicated manner, further the dependence is influenced by the distribution of the transfer times.  However one can derive the optimal policies by using an appropriate lower bound function. 

 We first consider the case:    \underline{$E[T] >  E[\Rcc]$, or    $d_n > d_o$}.  From (\ref{Eqn_Num_limit}) and the definitions following (\ref{Eqn_Num_limit}) and because of  positivity of the terms:  
  \begin{eqnarray}
\bar{a}(\theta)  \hspace{-2mm}& \ge & \hspace{-2mm}   f_o (\theta)  \mbox{ for any } \theta \ge 0,  \mbox{ with function, }  \nonumber \\  
f_o (\theta)  \hspace{-2mm} &:= &  \hspace{-2mm} \frac{ d_o  \big ( b_n\pi_\theta(0) +b_o \pi_\theta(1)  \big ) + 0.5( c_n \pi_\theta(0) +   c_o \pi_\theta(1))}{d_n \pi_\theta(0) +d_o \pi_\theta(1)}  \nonumber  \\
 \hspace{-2mm}& =&  \hspace{-4mm}\frac{  d_o  \big( (b_n - b_o )\pi_\theta(0)  +  b_o  \big ) + 0.5( c_n -c_o)  \pi_\theta(0) +  0.5 c_o  }{(d_n -d_o)\pi_\theta(0) + d_o }, \mbox{ with} \nonumber \\
              b_n &:=& E(T),  \  \  b_o := E(T | T<\xi)  = \frac{  E[T e^{-\lambda T}] }{ E[e^{-\lambda T}]}.
\end{eqnarray} 
Further using (\ref{Eqn_AAoI_DOP}) we have\footnote{It is not difficult to establish the continuity of the relevant functions as $\theta \to \infty$ and it is not difficult to show that the limit equals that with DOP scheme.} (e.g., $\pi_\theta(0) \to 0$ as $\theta \to\infty$):
\begin{eqnarray}
 \lim_{\theta \to \infty } f_o (\theta ) =\lim_{\theta \to \infty } {\bar a}(\theta) = {\bar a}_{DOP}.
 \label{Eqn_cmp_atheta_fl}
\end{eqnarray}
If  the  DOP scheme is optimal for the lower bound function 
$
f_o (\theta)  $,  i.e., if 
\begin{eqnarray}
\label{Eqn_DOP_opt_fl}
\min_{\theta } f_o  (\theta) = \lim_{\theta \to \infty}  f_o (\theta),
\end{eqnarray}
then DOP would be optimal for AAoI, because  then using (\ref{Eqn_cmp_atheta_fl}):
$$
{\bar a}_{DOP} \ge \min_{\theta } {\bar a}  (\theta) \ge \min_{\theta } f_o  (\theta) = 
\lim_{\theta \to \infty } f_o (\theta ) = {\bar a}_{DOP}.$$
We prove that (\ref{Eqn_DOP_opt_fl}) is true when DOP renewal cycle is smaller, and hence show the optimality of DOP (proof in Appendix A):   
  \begin{thm}
\label{Thm_dn_lt_do}
If $d_n \ge   d_o$ then DOP is optimal, i.e., 
$$
 \hspace{10mm} 
\min_{\theta  \ge 0} {\bar a}  (\theta) = \lim_{\theta \to \infty } {\bar a}(\theta) = {\bar a}_{DOP}.   \hspace{10mm} \blacksquare
$$
\end{thm}

It is clear from (\ref{Eqn_age_DNP}) and (\ref{Eqn_AAoI_DOP}) that the DOP scheme is better than the DNP scheme when its expected renewal cycle is smaller, i.e., when $d_n \ge d_o$.  Theorem \ref{Thm_dn_lt_do} proves much more under the same condition, the DOP scheme is better than any other threshold scheme. 
 
 We now study the reverse case, i.e., when $d_n < d_o$ or equivalently when  \underline{$E[T] <  E[\Rcc]$.}  In this case  $\bar{a}(\theta)  > f_n (\theta)$ where
  \begin{eqnarray*}
  f_n (\theta) :=  \frac{ d_n ( b_n\pi(0) +b_o \pi(1) ) + 0.5 c_n \pi(0) + 0.5 c_o \pi(1)}{d_n \pi(0) +d_o \pi(1)} .
\end{eqnarray*}
As in the previous case, if DNP  is proved optimal  for this lower bound function,  then  DNP is optimal for controlled AAoI, and this is proved in the following (proof in Appendix A): 
\begin{thm}
If $d_n <  d_o$ and  (note that\footnote{
because
 $$1- \lambda E\left [T e^{-\lambda T  } \right ]  
 = \lambda    ( E[ \xi  ] -  E[ T ;  T\le \xi ] )  = \lambda  (E[ \xi ; T > \xi  ]  + E[ \xi - T ;  T\le \xi ])  > 0.
 $$
} $1- \lambda E\left [T e^{-\lambda T  } \right ] > 0$)
\begin{eqnarray}
\label{Eqn_DNP_opt_condition}
\rho  \frac{E[T^2]}{2 E[T]}    - (1+\rho)   \left ( d_o    -d_n  \right ) \left ( 1- \lambda E\left [T e^{-\lambda T  } \right ]  \right ) 
\le 0,
\end{eqnarray} 
then DNP is optimal,  \vspace{-5.9mm}
$$
\hspace{18mm} 
\min_{\theta \ge 0}  {\bar a} (\theta) =  {\bar a} (0) = {\bar a}_{DNP}.  \hspace{10mm} \blacksquare
$$
  \label{Thm_dn_gt_do}
\end{thm}

\subsection{Stationary  Markov Randomized policies} 
\label{sec_SMR}
We now  generalize the results of previous subsection by considering  Stationary Markov Randomized (SMR) policies. As seen from (\ref{Eqn_SLLN_reqd2}) the objective function AAoI 
is the ratio of two average costs and hence the  usual techniques of Markov decision processes may not be applicable. Nevertheless we could use exactly  the same techniques as in previous subsection to show the optimality of DNP/DOP policy even under SMR policies. This is true  under the assumptions of Theorems \ref{Thm_dn_lt_do}-\ref{Thm_dn_gt_do}.

Let $\Policy^\infty$ be any Stationary Markov randomized   policy:   $\Policy (G)$ represents the probability with which DNP scheme is selected when the state $G_k = G$, and this is true for any decision epoch $k$. Re define  $X_k = 1$ if DOP scheme is selected, else $X_k = 0$.  Like before, the random variables $X_k, {R_c}_k$ and $G_k$ depend mainly upon $X_{k-1}$, and same is the case with their stationary distributions. Let $\pi_\alpha$ represent the stationary  probability  that  $\{X^* = 0\}$, when policy $\Policy^\infty$ is used and note that:
\begin{eqnarray}
\pi_\alpha  &:=& \pi_\alpha (0) = \frac{q_\alpha}{q_\alpha + p_\alpha}  \mbox{ with }  \\
q_\alpha &:=&  E[ \Policy (\underline{T}) ]= E[  \Policy  (T) | T \le  \xi ]  \mbox{ and } p_\alpha =  E[1-  \Policy ( T) ]. \nonumber
\end{eqnarray}
As before,
the stationary expectation   (see  (\ref{Eqn_SLLN_reqd}))
\begin{eqnarray*}
E_{\pi_\alpha} [ G_{l-1} {R_c}_l ] \hspace{-2mm} &\hspace{-1mm}=\hspace{-1mm} &  \hspace{-2mm}  E_{\pi_\alpha} [ G_{l-1} {R_c}_l ; X_{l-1} = 1 ]  +  E_{\pi_\alpha} [ G_{l-1} {R_c}_l ; X_{l-1} = 0 ]   \nonumber 
\\
&& \hspace{-18mm} = \   
d_n  E[G_*  E[ X^* = 1 | G*] ]   +  d_o E[G_*  E[ X^* = 0 | G_*]  ]  \\
&& \hspace{-18mm} = \   
d_n  E[G_*   \Policy (G*) ]   +  d_o E[G_*  (1-  \Policy (G*) ) ]  \\  .
&& \hspace{-18mm} = \   
d_n   \Big ( E[T   \Policy (T) ] \pi_\alpha  (1) +   E [ T \Policy (T) |  T \le \xi ] \pi_\alpha  (0 )  \Big )   \\
&& \hspace{-13mm}  + \   
d_o  \Big  ( E[T  (1- \Policy (T) )] \pi_\alpha  (1) +   E [ T (1- \Policy (T) )  |  T \le \xi ] \pi_\alpha  (0 )  \Big ) .
\end{eqnarray*}
Similarly
\begin{eqnarray*}
E_{\pi_\alpha} [{R_c}_ k^2] &=&   E_{\pi_\alpha} [ {R_c}_ k^2 ; X_{k-1} = 0]  +  E_{\pi_\alpha} [ {R_c}_ k^2 ; X_{k-1} = 1]  \\
&=&  c_o  \pi_\alpha (1) + c_n \pi_\alpha   (0) ,
\end{eqnarray*}
Proceeding  exactly as in the case of threshold policies:
\begin{eqnarray*}
{\bar a} (\Policy)  &=&   \frac{\beta_\alpha (0)  \pi_\alpha (0) + \beta_\alpha  (1)  \pi_\alpha (1)}{ d_n \pi_\alpha (0) +  d_o \pi_\alpha (1) } \mbox{ with } \\
\beta_\alpha (0) &:=& d_n E [ T \Policy (T) ] +  d_o  E [ T (1- \Policy (T) )   ] + 0.5 c_n  \mbox { and }\\
\beta_\alpha (1) &:=& d_n E [ T \Policy (T) |  T \le \xi ] +  d_o  E [ T (1- \Policy (T) )  |  T \le \xi ] + 0.5 c_o .
\end{eqnarray*} 
Using the  lower bound functions, $f_o (\cdot)$ and $f_n (\cdot)$, and following exactly the same logic  one can extend Theorems \ref{Thm_dn_lt_do}-\ref{Thm_dn_gt_do}: 
\begin{thm}
\label{Thm_Markov_policies}
a) If $d_n \ge   d_o$ then DOP is optimal among SMR policies, i.e.,  \vspace{-4mm}
$$
 \hspace{10mm} 
\min_{\Policy^\infty \in  SMR} {\bar a}  (\alpha) =   {\bar a}_{DOP}.   
$$
b) If $d_n <  d_o$ and  (\ref{Eqn_DNP_opt_condition}) of Theorem \ref{Thm_dn_gt_do} is true
then DNP is optimal,  \vspace{-.5mm}
$$
\hspace{18mm} 
\min_{\Policy^\infty \in  SMR} {\bar a}  (\alpha) =  {\bar a}_{DNP}.  \hspace{10mm} \blacksquare
$$
\end{thm}
\begin{table}[h]
\centering
\bgroup
\def\arraystretch{0.20}

\caption{Criterion for ${\bar a}_{DOP } \le  {\bar a}_{DNP}$, for different types of $T$ }
\label{Table_DOP optimal}
\hspace{-6mm}
{\small
\begin{tabular}{|c|c|c|}
\hline  && \\  
Distribution & CDF ($ P(T \leqslant x)$)  &  ${\bar a}_{DOP } \le  {\bar a}_{DNP}$ when \\ && \\  \hline  && \\
 Uniform (0,$\phi$) & $ \Big[\frac{x}{\phi}\Big ] 1_{  x > 0}$ &    \hspace{-2mm}$
\frac{1}{ 1 - e^{-\lambda \phi }  } < \left ( \frac{1}{3  }  \frac{ \lambda \phi }{2 + \lambda \phi } + \frac{1}{\lambda \phi } + \frac{1}{2}  \right ) 
$ 
\\

&& Approximately $\lambda \phi <
2.356$   \\ &&
\\ \hline  && \\
Weibull ($\mu$, k) & $\Big[1- e^{ -(x/ \mu)^k}\Big ] 1_{ x > 0} $  \hspace{-3mm} &  \hspace{-2mm}  $      \frac{\rho^2  w_{k}^2  }{2(1+\rho w_{k}^1) } +  (1 + \rho w_{k}^1 )    > \frac{1}{w_{\rho}^k} $ \hspace{-4mm}
 \\ 
&&   $w_{k}^i = \Gamma( 1+ \frac{i}{k})$   \&  $\rho =\lambda \mu$ 
\\
&&
$w_{\rho}^k  = E_k [  e^{ -\rho T}  ]$ \\ &&
\\ \hline
 Exponential ($\mu$)& $ \Big[1- e^{-\mu x}\Big ]  1_{  x > 0}$ &  all $\mu$ \\ \hline
 Hyperexpo($\{\mu_i , p_i\}_i$)\hspace{-2mm} & $\Big[1- \hspace{-2mm}\sum\limits_{i=1}^{n}p_i e^{-\frac{x}{ \mu_i}}\Big ] 1_{  x > 0} $& all $\{\mu_i , p_i\}_i$\\ \hline
\end{tabular}}
\egroup
\end{table}

\subsection{Numerically aided study}
Theorems \ref{Thm_dn_lt_do} and \ref{Thm_Markov_policies} show that the  DOP scheme is optimal, when the expected renewal cycle with DOP scheme ($d_o$)  is smaller than that ($d_n$) with DNP. 
While Theorems \ref{Thm_dn_gt_do}-\ref{Thm_Markov_policies}, show the DNP scheme is optimal under converse conditions. However the optimality of DNP scheme requires an extra condition (\ref{Eqn_DNP_opt_condition}).  

\subsubsection*{DOP optimal among SMR policies} We considered several distributions for transfer times and tested the conditions required for DOP/DNP optimality.
The results are summarized in Table \ref{Table_DOP optimal}. 
 By direct substitution one can show that $d_n = d_o$ for exponential and $d_n < d_o$ for hyper exponential distribution.  Thus by Theorem \ref{Thm_Markov_policies}, DOP is optimal for these transfer times.

One can similarly derive the conditions under which DNP is optimal among SMR policies, given the distribution of transfer times.

\begin{figure}
\vspace{-15mm}
\includegraphics[scale=.3]{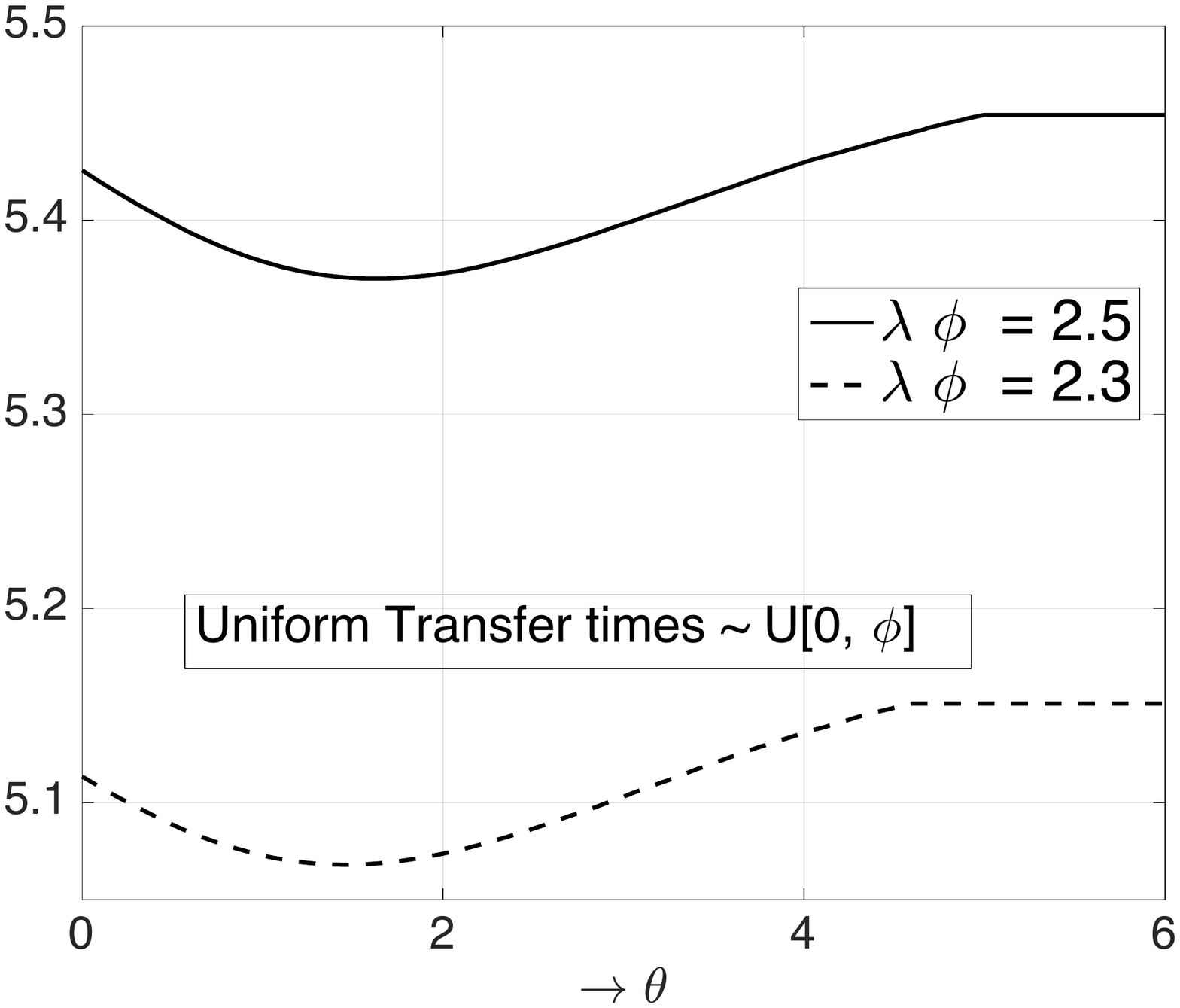}
\vspace{-24mm}
\caption{ AAoI versus $\theta$ for uniform transfer times: Intermediate $\theta$ optimal but DNP/DOP almost optimal \label{fig_inter}}
\end{figure}
\begin{figure}
\vspace{-20mm}
\includegraphics[scale=.3]{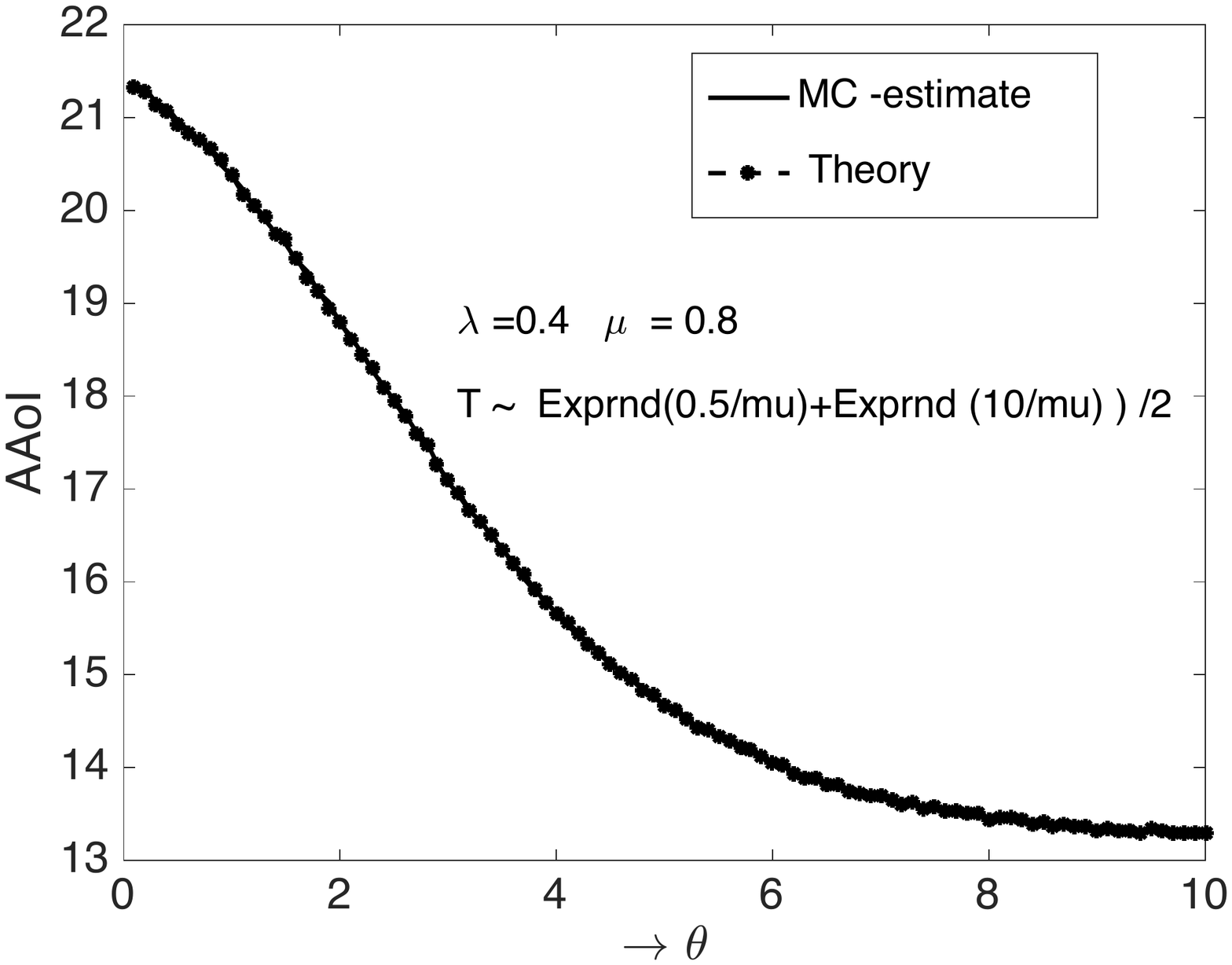}
\vspace{-25mm}
\caption {When $d_o = 13.19 >  d_n = 12.97$ and  bound  (\ref{Eqn_DNP_opt_condition}) 
 > 0:   optimizer is DOP  \label{Fig_DNP_Conditions_voilate}}
\end{figure}

\subsubsection*{DNP/DOP is almost optimal}
When $d_n < d_o$, but (\ref{Eqn_DNP_opt_condition}) is not satisfied,  we do not have  theoretical understanding of the optimal policy.  We study such test cases by numerically optimizing  (\ref{Eqn_AAoI_theta}) over threshold policies.
%
One such example is plotted in Figure \ref{Fig_DNP_Conditions_voilate}, which considers Erlang distributed transfer times.  The AAoI is plotted as a function of $\theta$, it decreases as $\theta \to \infty$, hence confirming that the AAoI is minimized by DOP scheme.  


A second  example is considered in Figure \ref{fig_inter}  with uniformly distributed transfer times.  Here  again  AAoI ${\bar a} (\theta)$ is plotted as a function of $\theta$ for two different parameters.  
An intermediate $\theta \in (0, \infty)$ is   optimal in both the  examples of this figure,  however  DOP and DNP perform almost similar. Further   AAoI at  $\theta^*$ is close to that at DNP/DOP
 (Figure \ref{fig_inter}).  We considered many more such case studies and observed similar pattern: DOP/DNP scheme is (almost) optimal. 
 These examples include  truncated exponential, Log normal, Poisson distributed  and Erlang transfer times  etc.

\subsubsection*{Best among DNP/DOP} 
Thus  either DNP or DOP scheme is (almost)  optimal among the threshold policies. Hence it is important to derive the conditions that suggest the best among the two. 
One can find the best among DNP/DOP schemes   by directly  using (\ref{Eqn_age_DNP}) and (\ref{Eqn_AAoI_DOP}), i.e., DNP  is better than DOP
if  and only if  (iff)
\begin{eqnarray}
E[T] -  \frac{1-\gamma }{\lambda \gamma }  +  \frac{E[T^2]}{2 E[T]}   \frac{\rho}{1+ \rho}   < 0 \mbox{ or iff  } \nonumber  \\
d_o - d_n  > \frac{E[T^2]}{2 E[T]}   \frac{\rho}{1+ \rho} \mbox{ or iff }  \nonumber \\
 1   >  \left (  \frac{E[T^2]}{2 (E[T])^2 }   \frac{ \rho^2 }{1+ \rho} + 1+ \rho \right ) \gamma .
\label{Eqn_DNP_opt_overl_condition}
\end{eqnarray}
Note that $E[T^2] = variance (T) + (E[T])^2$ and we have the following important conclusions:
\begin{itemize}
\item DOP is the best  for small update rates:   as the update rate $\lambda \to 0$, with the distribution of $T$ fixed,  the above condition  is negated (RHS is approximately $1+\rho$) . 
\item DNP is the best  for large update rates:   as the update rate  $\lambda \to 0$, with distribution of $T$ fixed,  the above condition is satisfied (RHS converges to 0).
\item The range of $\lambda$ for which DNP is optimal is influenced by the variance.   {\it 'DOP scheme becomes optimal as the variance  of the transfer times  increases, for  bigger range of $\lambda$'.} 
\end{itemize} 
For uniform transfer times   we derived the conditions   under which DOP performs better than DNP, using (\ref{Eqn_DNP_opt_overl_condition}), and the condition is tabulated in
the first row of Table \ref{Table_DOP optimal}.  Approximately, DOP is optimal if  $\lambda \phi < 2.35$.  
Weibull is also tabulated in the second row.

 Based on this theoretical and numerical case studies we have the following overall important conclusions 
\begin{itemize}
\item AAoI is  (almost) optimized either by DOP scheme or by DNP scheme. No  other threshold  policy performs significantly better  than the best among these two static policies.
\item If  expected renewal cycle with DOP is smaller than that with DNP,  DOP scheme  optimizes  AAoI  over all SMR policies. 
\item  When DNP has smaller renewal cycle,  DOP may  still be the optimal (in some test cases). 
\end{itemize}

The Figure \ref{Fig_DNP_Conditions_voilate} also plots the Monte-Carlo  estimates of AAoI along with formula (\ref{Eqn_AAoI_theta}).  For Monte-Carlo estimates we generate several random sample paths
and compute the time average of AAoI.  As anticipated,  the formula well matches the estimates (see Figure \ref{Fig_DNP_Conditions_voilate}).


\Detail{
The above table \ref{DOP optimal} shows the  different distribution cumulative  distribution function and last column shows under what condition DOP is optimal

\subsection{Examples and DNP-DOP comparison}
For exponential transmission times,  i.e., when $T$ is exponentially distributed with parameter $\mu$,  
$$
\gamma = \frac {\mu } {\lambda + \mu} \mbox{ and }  (\rho+1) \gamma =  \left ( \frac{\lambda}{\mu} + 1 \right  ) \gamma = 1.
$$
Thus by  Theorem \ref{Thm_dn_lt_do}, DOP is optimal.

As already observed  from simulations and theoretical results, either DNP or DOP (and no other Markov policy) is almost optimal.  
Thus for general transition times,  
  one can  find the best  directly using (\ref{Eqn_age_DNP}) and (\ref{Eqn_AAoI_DOP}), i.e., the optimal scheme is DNP 
if  (and only if)
\begin{eqnarray}
E[T] -  \frac{1-\gamma }{\lambda \gamma }  +  \frac{E[T^2]}{2 E[T]}   \frac{\rho}{1+ \rho}   < 0 \mbox{ or if  } \nonumber  \\
d_o - d_n  > \frac{E[T^2]}{2 E[T]}   \frac{\rho}{1+ \rho} \mbox{ or if }   1   >  \left (  \frac{E[T^2]}{2 (E[T])^2 }   \frac{ \rho^2 }{1+ \rho} + 1+ \rho \right ) \gamma .
\label{Eqn_DNP_opt_overl_condition}
\end{eqnarray}
Note that $E[T^2] = variance (T) + (E[T])^2$ and we have the following important conclusions:
\begin{itemize}
\item DOP is optimal for small update rates:   as the update rate $\lambda \to 0$, with distribution of $T$ fixed,  the above condition  is negated (RHS is approximately $1+\rho$) . 
\item DNP is optimal for large update rates:   as the update rate  $\lambda \to 0$, with distribution of $T$ fixed,  the above condition is satisfied (RHS converges to 0).
\item The range of $\lambda$ for which DNP is optimal is influenced by the variance.   {\it 'DOP scheme becomes optimal as the variance  of the transfer times  increases, for  bigger range of $\lambda$'.} 
\end{itemize}

For example,  when the transition times are  \underline{uniformly distributed} between $0$ and $\phi$,  DNP is optimal if and only if:  
$$
 1 \ge  \left ( \frac{1}{3  }  \frac{ \lambda \phi }{2 + \lambda \phi } + \frac{1}{\lambda \phi } + \frac{1}{2}  \right )  ( 1 - e^{-\lambda \phi }  )
$$
When $\lambda \phi$ is small (as $ 1 - e^{-\lambda \phi }   \approx \lambda \phi $), DOP is optimal.  On the other hand for large values of $\lambda \phi $ (the RHS converges to zero)
DNP is optimal. By numerical computations, we observe that for  $\lambda \phi < 2.356$ DOP is optimal, while DNP is optimal otherwise.

\underline{Weibull Distribution:} Let $T_\mu$ follow  Weibull distribution with scale parameter $\mu$ and the shape parameter $k$:
\begin{eqnarray}
\label{Eqn_CDF_Weibull}
F_{\mu, k} (x) = Prob ( T_\mu \le x) = 1 - e^{ - (x/ \mu)^k}  \mbox{ for all }  x \ge 0.
\end{eqnarray}

Consider $T=T_1$, with scale parameter 1 and define the following:
 \begin{eqnarray*}
 w_{k,1}  &:=& E(T)  =   \Gamma(1+\frac{1}{k}) ,\  \   \   w_{k,2}  \ := \  E(T^2) =     \Gamma(1+\frac{2}{k})  \mbox { and }  \\
\gamma_k  &= &  E_k[e^{-\lambda T} ]= \sum_{n=0}^{\infty} \frac{(-1)^n(\lambda )^n  \Gamma(1+\frac{n}{k})}{n!}  .
 \end{eqnarray*}
 
 From (\ref{Eqn_CDF_Weibull}),    $T_\mu \stackrel {d}{=} \mu T$. Thus
one can rewrite $\gamma_{\mu, k} = E_{k}[ e^{- \rho T}]$, where $T$ is Weibull distributed with spread parameter $1$ and $\rho = \mu \lambda$. 
Then the condition  for DNP optimality is  (from (\ref{Eqn_DNP_opt_overl_condition}))
\begin{eqnarray}
\left ( \frac{ w_{k,2} }{2}      \frac{\rho^2    }{1+\rho w_{k,1} } +  (1 + \rho w_{k,1} )   \right ) E_k [  e^{ -\rho T}  ] < 1.  
\end{eqnarray}
Thus for any fixed $k$, the failure parameter, there exists a range of $\rho$, for which DNP is optimal and DOP is optimal otherwise. As $\rho$ increases to $\infty$, the LHS decreases and hence DNP would be optimal for all $\rho \ge {\bar \rho}$, where the upper bound ${\bar \rho} $ will depend upon $k$. 
 
 On the other hand, when $\rho$ is fixed and when $k \to \infty$ then (using  dominated convergence theorem) one can show that 
 $$
 \gamma_k = E_k [e^{-\rho T}]  \to  e^{-\rho}, \mbox{ and } c_{k,1} \to 1,  c_{k,2} \to 1.
 $$
 Thus DNP is optimal (asymptotically in $k$) if 
 $$
 \left (      \frac{\rho^2    }{1+\rho  } +  (1 + \rho   )   \right )    e^{ -\rho  }    <  1,
 $$else DOP is better than DNP.
 Thus for large $k$, DNP is optimal if load factor is large. If load factor $\rho$ is small, then DOP is optimal. 

Thus optimality of DNP/DOP depends upon both the failure rate co-efficient $k$, as well as $\rho$ (representative of load factor) which in turn depends upon the spread of the distribution. 

\textcolor{red}{
Theoretically DOP is optimal if  $d_o - d_n \le 1$, or if
\begin{eqnarray}
E_k  [e^{-\rho T} ]  \left  (  1  + \rho w_{k,1}  \right )  \ge 1.
\end{eqnarray}}

\underline{Hyper Exponential:} Let T be the hyper exponential distribution with parameter $p_i$ and $\mu_i$. Then
 \begin{eqnarray*}
 E(T)= \sum_{i=1}^{n}\frac{p_i}{\mu_{i}  }   \mbox{ and }    \gamma = \sum_{i=1}^{n}\frac{\mu_{i}p_i}{(\mu_{i} +\lambda)} 
 \end{eqnarray*}
 By direct comparison $ d_ n > d_ o $, i.e.,    DOP is  optimal,  if the following holds:
 \begin{eqnarray*}
  1+ \lambda  \sum_{i=1}^{n}\frac{p_i}{\mu_{i}  }  > \sum_{i=1}^{n}\frac{\mu_{i}p_i}{(\mu_{i} +\lambda)}
  = \sum_{i=1}^{n} p_i   \left  ( 1 -  \frac{\lambda }{(\mu_{i} +\lambda)} \right ) = 1
 \end{eqnarray*}
  But this is always true as:
 \begin{eqnarray*}
 \sum_{i=1}^{n}\frac{\mu_{i}p_i}{(\mu_{i} +\lambda)}
 & =&  \sum_{i=1}^{n} p_i   \left  ( 1 -  \frac{\lambda }{(\mu_{i} +\lambda)} \right ) 
  \  = \  1  - \lambda  \sum_{i=1}^{n}\frac{p_i}{\mu_{i} + \lambda }\\
  & <&  1   <  1+ \lambda  \sum_{i=1}^{n}\frac{p_i}{\mu_{i}  } .
 \end{eqnarray*}
 
  Thus DOP is always optimal for Hyper exponential distribution.

\textcolor{red}{ For which distributions DOP is optimal condition \ref{Eqn_DNP_opt_condition} negated/satisfied.  I noticed it for  $\log (1 +  h^2 / sig^2) $, where $h^2$ is exponentially distributed. We can make a table of the distribution and the ranges/parameter ranges for which DOP/DNP is optimal.  }

\subsection*{Decision epochs}
We consider the successful message delivery epochs as the decision epochs. It may be more natural to decide if a new arrival should be dropped or 
accepted while dropping the old one (under transmission) at its arrival instance itself. That is, it may be more natural to  consider the message arrival epochs as the decision epochs. We instead considered message delivery epochs as the decision epochs,  for mathematical tractability.  However the results are still interesting  because
DOP/DNP  renewal cycles have at maximum 2 to 3 arrivals, which implies we are taking decisions at once for around 2 to 3 arrivals.  In future we may consider the other decision epochs. 

\textcolor{red}{ We can demonstrate through MC simulations that threshold schedulers at every arrival epoch are also optimized by DOP or DNP policy. Also discuss that this cost is ratio of two average costs and hence complicated, in case we succeed with establishing optimality w.r.t. Markov randomized policies. }

}{}

 \section{Multiple sources} 
 \label{sec_multiple}

  In   previous sections single source problem is considered and optimality of DNP/DOP scheme is established.  We now consider multiple sources  ($S$ number of them) transmitting information over a common channel to a common destination.
 We again consider the case with zero storage and derive the performance of the two schemes.   The sources are assumed to have capability to detect the silence before they commence  transfer of  their information.  
Basically they use Carrier-sense multiple access (CSMA) protocol to detect the silence. No source can  interrupt the transmission of the packets from the other source. 
   Because of zero storage,  arriving packets  of a source are dropped, if any other  source is transmitting.

\subsection{ DNP scheme}
 
    We begin with DNP protocol, where the new packets of the transmitting source  are also  dropped. 
    In this case, one needs to consider separate  renewal cycles  for each source.  A $k$-th renewal cycle for source $s$, is the time duration between $(k-1)$-th and $k$-th successful packet reception at destination from source $s$.  As in single source DNP scheme, these renewal cycles are identically distributed, but are not independent. However once again the alternate (even and odd) cycles are independent and hence RRT can be applied.  Thus formula (\ref{Eqn_AAoI_by_RT_limit}) is applicable, and we need to compute the first two moments of the renewal cycles corresponding to each source and the age at the beginning of the  source-wise renewal cycles.

    Let $\{R_{s,k} \}_k$ be the renewal epochs  corresponding to source $s$.  
    Any renewal cycle starts with an idle period which is exponentially distributed  with parameter  $\lamS := \sum_{s \le S} \lambda_s$.  Basically this is the time at which an update packet   is available at one  of the sources.  This is  followed  by  a `busy period' during which many messages (of other sources) are transferred with phases of silences in between and finally ends with transfer of message of source $s$.  
%
%
%
%
%
Thus a  renewal cycle of source $s$ consists of ${\mathcal N}_s$, the  Geometric  number (with parameter $\lambda_s /  \lamS$)  of 
sub-cycles in which the other sources transfer, followed by the last one in which source $s$ transfers.   In between such transfer periods, we have one idle period which is exponentially distributed with parameter $\lamS$.
Thus the renewal cycle corresponding to source $s$ can be expressed\footnote{
To simplify the notations, we are suppressing the cycle number $k$  while representing the random quantities.  Here $i$ represents the sub-cycle number.} as the following:
\begin{eqnarray}
\label{Eqn_Rc_multiple_source}
{R_c}_{s,k} 
 =   \sum_{i=1}^{{\mathcal N}_{s, k}-1 }  \sum_{s' \ne s} \left  (  {\underline \IA}_{s', i} + T_{s', i}  \right ) {\mathcal I}_{s', i} +  {\underline \IA}_{s, {\mathcal N}_{s, k}} + T_{s,{\mathcal N}_{s, k}}, 
\end{eqnarray} where ${\mathcal I}_{s',i}$ is the indicator that the packet of source  $s'$ arrived before that of other sources during the $i$-th idle period  and
 ${\underline \IA}_{s', i} $  is the `special inter arrival time' of source  $s'$  given that it arrived before that of the other sources (during $i$-th idle period and within the $k$-th renewal cycle):
\begin{eqnarray}
\label{Eqn_special_xi}
{\underline \IA}_{s',i} := \IA_{s',i} \bigg  |   \bigg \{ \IA_{s',i} \le \IA_{s'',i} ;  \mbox{ for all }  s'' \ne s' \bigg  \}.
\end{eqnarray}
Further unlike in the single source case, the  transfer time ($ T_{s,{\mathcal N}_{s, k}}$)  is  not special, but it is distributed like any typical transfer time $T_s$. 
Thus  $E[G_s] = E[T_s]$.

Conditioning as before, and with $ {\tilde  {R_c}}_s  $ representing an IID copy of renewal cycle of source $s$:
\begin{eqnarray*}
E[{R_c}_s ] &=&    \sum_{s' \ne s}  E\left [{R_c}_{s'} ;    {\mathcal I}_{s',1} \right   ]    \\ 
&=& \sum_{s \ne s} E \left [ ( {\underline \IA}_{s',1} + T_{s',1}) +  {\tilde  {R_c}}_s ;     {\mathcal I}_{s',1}  \right ]
 +   E \left [ {\underline \IA}_{s,1}  + T_{s,1}   ;     {\mathcal I}_{s,1} \right  ]
\end{eqnarray*}
Thus simplifying (note $E[ {\mathcal I}_{s,1} ] = \lambda_s  /\lamS$ and  $E[R_c] = E[{\tilde R}_c]$)
\begin{eqnarray}
E[{R_c}_s ] &=& \frac{ \sum_{s' }    E \left [ {\underline \IA}_{s',1} + T_{s',1}  ; I_{s', 1}  \right ] } {   \lambda_s } \Detail{\mbox{ with }
\\
\nonumber 
E[  {\underline \IA_s} ]  &=& E[ \IA_s  | \IA_s <  \min \{ \IA_{s'} \}_{ s \ne s' } ]   \nonumber  \ = \  \Detail{\frac{  E[  \IA_s ;  \IA_s <  \min \{ \IA_{s'} \}_{ s \ne s' } ]}{P(\IA_s <    \min \{ \IA_{s'} \}_{ s \ne s' })}   \nonumber \\
 &=&
 \frac{\lamS }{ \lambda_s} \frac {\lambda_s } {  \lamS^2 } =  }{} \frac{1} {\lamS }  .}{}
 =  \frac{1 +  \sum_{s'=1}^S  \lambda_{s'} E[T_{s'}]  } {\lambda_s}.  \hspace{9mm} 
 \label{Eqn_Rc_Multiple}
\end{eqnarray}
In a similar way
\begin{eqnarray}
E[{R_c}_s^2 ] & =&  \frac{\sum_{s'} E [ ( \xi_{s'} + T_{s'} )^2  \mathcal{I}_{s'} ] + 2 E[{R_c}_s] \sum_{s' \ne s}   E[ (\xi_{s'} + T_{s'} ) \mathcal{I}_{s'} ]}
{\lambda_s} \nonumber  \\
& = &
 \frac{1}{\lambda_s}   
\Bigg [   \frac{2 }{\lamS}  + \sum_{s'}  \lambda_{s'} E[T_{s'}^2]    + 2 \sum_{s'}  \frac{ \lambda_{s'} E[T_{s'}] }{\lamS}  
\nonumber  \\
&& \hspace{13mm}
+ 2 E[{R_c}_s]   \left (   \sum_{ s' \ne s}   \lambda_{s'} + \sum_{ s' \ne s}    \lambda_{s'} E[T_{s'}]  \right ) 
\Bigg  ]. \label{Eqn_Rc2_Multiple}
\end{eqnarray}
Substituting the above into (\ref{Eqn_AAoI_by_RT_limit})    one can compute the AAoI ${\bar a}_{DNP, s}$. 

\ignore{

\begin{eqnarray*}
E[({R_c}_s)^2 ] &=&    \sum_{s' \ne s}  E\left [( {R_c}_{s'} )^2 ;    {\mathcal I}_{s',1} \right   ]    \\ 
&=& \sum_{s \ne s} E \left [ ( ( {\underline \IA}_{s',1} + T_{s',1}) +  {\tilde  {R_c}}_s  )^2;     {\mathcal I}_{s',1}  \right ] \\
&& +   E \left [ ( {\underline \IA}_{s,1}  + T_{s,1}  )^2  ;     {\mathcal I}_{s,1} \right  ]  \\
 &=& \sum_{s \ne s} E \left [ ( ( {\underline \IA}_{s',1} + T_{s',1})  )^2  ;     {\mathcal I}_{s',1}  \right ] \\
 &&+  \sum_{s \ne s} E \left [ ( {\tilde  {R_c}}_s  )^2     \right ]  E\left [     {\mathcal I}_{s',1}  \right ] \\
 &&+  2 E \left [   {\tilde  {R_c}}_s \right ]  \sum_{s \ne s} E \left [     {\underline \IA}_{s',1} + T_{s',1})       ;     {\mathcal I}_{s',1}  \right ] \\
 &&+   E \left [ ( {\underline \IA}_{s,1}  + T_{s,1}  )^2  ;     {\mathcal I}_{s,1} \right  ]
\end{eqnarray*}}

\subsubsection*{DOP scheme}
One can  alternatively   consider DOP scheme.
None of the sources can interrupt the transmission of the other source. However  the transmitting source  will drop the old packet, if   interrupted by a new arrival of the same source. 
The analysis will be exactly similar and one can derive the AAoI of each source with DOP scheme by replacing $E[T_s]$ and $E[T_s^2]$ respectively with $E[\Rcc_s]$ and $E[\Rcc_s^2]$
in equations (\ref{Eqn_Rc_Multiple})-(\ref{Eqn_Rc2_Multiple}) and by replacing $E[G] = E[\Rcc]$.

One can again compare the two schemes, which clearly depends upon the values of $E[T_s]$- $E[T_s^2]$ and  $E[\Rcc_s]$- $E[\Rcc_s^2]$ for all sources. However we prefer to discuss a more practical scenario with multiple sources, that with single storage.  We may consider this topic  in future.

With multiple sources there is an inherent competition, each source would  like to reduce their individual AAoI.  Towards this, each source might want to chose appropriate update rate. 
This is a very {\it important aspect in analysis related to the Freshness of information (\cite{HowOften} etc).}    Most of the work concentrates on optimal update rates $\{\lambda_s\}$.  Using the performance analysis obtained in this paper one can {\it also discuss optimal update rate in  case of single source, and an appropriate game theoretic formulation in the case of multiple sources}.  This aspect could be of future interest. 
However our focus in this paper has been on a completely different aspect, the  correct packets to be dropped.  
  We continue with the same subject.

\ignore{
\subsection{DOP scheme} 
Once again no source can interrupt the transmission of the other source. However  the transmitting source  will drop the old packet, if   interrupted by a new arrival of the same source.  
 
 Analysis:  the only difference is in the renewal cycle.  We will have
 \begin{eqnarray}
E[{R_c}_1 ] &=& \frac{ \sum_{s=1}^S  \lambda_s  E \left [ {\underline \IA}_{s,1} + {\Rcc}_{s,1})    \right ] } { \lamS -  \lambda_1 } \mbox{ with }
\end{eqnarray}
.
.
.
. 
 
\subsection{Comparison of the two schemes}}

\section {Multiple sources with one storage}
\label{sec_multiple_storage}

In multiple source scenarios,  as the number of sources increase, the transmission chances  per source gets reduced.  In such heavy traffic/congested scenarios, it  is inefficient to wait for the new packets.  Hence we consider storage possibilities at the source.  It is {\it sufficient to consider one storage at every source, as the old packet gets obsolete once a new packet arrives.}   The waiting packet (if any) is  replaced by the new arrival (of the same source).   Once again with more number of sources sharing the common channel, and with rare transmission opportunities per source, one can assume that the sources always  have a  packet to transmit at the moment the channel becomes available/silent. This is like the well known saturation case (e.g., \cite{slot}). Thus one needs a contention resolution   algorithm, which is preferably  decentralized.  We consider a collision based contention resolution protocol (like slotted Aloha protocol) when the transmission is over    random channels, in particular Rayleigh channels. 

We establish an interesting connection between Rayleigh/random channels and DOP scheme.  The  DOP protocol, when optimal, enjoys supremacy because of a \underline{`reverse  bus paradox'}.  In DOP protocol a old packet is dropped, if its transmission is not completed before the arrival of the new packet.  Thus there is a natural rejection/dropping of `bad' (long transfer time) packets, which results in optimality of DOP protocol for the cases with large variability as discussed in  previous sections. 
When multiple sources attempt to transfer their message over (independent) random channels,  there will be a similar natural rejection/dropping of the `bad' packets (to be more precise the packets that arrived during bad channel conditions). Thus we will observe that  the presence of multiple sources actually enhances the performance of each source. 

\subsection{CSMA Protocol with storage} 
Consider  again $S$ sources  competing to transfer their messages regularly to a common destination.    
As already mentioned all the sources are saturated, i.e., have an update packet to transfer at the moment the channel gets silent. 
Each source attempts transmission independent of the other sources, source $s$ attempts with  probability $q_s$.  
The source(s) first send control packets, with a  request to transmit.  If  the destination receives at least one of such requests successfully, it allocates the  channel to the owner of one among the successfully received  requests.  The selected agent transfers its packet via the channel.

We assume quasi static channels:  channel quality factors of each attempt (control request plus data transfer phase),  corresponding to any agent, are assumed to be IID. 
Let $H_s$  represent  the channel gain  of   agent $s$  and let $\At_s$ represent the flag that agent $s$ attempted transmission (note   $E[\At_s]=q_s$)
in the current attempt chance.  Then the   (control) Signal to Noise plus Interference Ratio (SINR) of agent $s$ is given by:
$$
\Phi_s =  \frac{ |H_s |^2 \At_s }{ \sum_{s' \ne s} |H_{s'}|^2 \At_{s'} + \sigma_n^2 },
$$where $\sigma_n^2$ is the noise variance. 
If many sources attempt together there is a collision (control SINRs are weak), but if  few source(s) attempt,  
control packet(s)  may reach the destination successfully.
 If the  received (control) SINR is above a threshold $\theta$, i.e., if   $$\Phi^* := \max_s \Phi_s  >  \theta, $$  
the  control packet(s) reaches the destination successfully  and the destination accepts the one with the best SINR.  And then the transmission of the actual packet happens at rate inversely proportional to $\log(1+SINR)$, but now the SINR corresponds to the case when the selected source transmits alone. Thus the time taken to complete the transmission (if  $  \Phi^* $ is greater than $\theta$) equals:
\begin{eqnarray}
T^* &=& T_{s^*} \mbox{ with } T_s := \frac{C}{\log (1 +   \Psi_{s} )      } ,   \ 
\Psi_s = \frac{ |H_s|^2   }{  \sigma_n^2 } \mbox{ for any } s \mbox { and } \nonumber \\
  s^* &\in& \arg \max_s \{\Phi_s \}.
\end{eqnarray}
In the above $C$ is an appropriate constant. 
Recall the channel strengths $\{H_{s, k} \}_k$ across various time slots (for each $s$) are IID.

Any time a source transfers the fresh message. 
So, once the destination receives a message from the source $s$ the age of the corresponding information reduces\footnote{Given a packet is in buffer, the age of the packet  is exponentially distributed with the same parameter $\lambda$.} to $T_s + \xi_s$. {\it Without loss of generality normalize the length of control phase to one.} Thus the age of any agent modifies in the following manner at the end of  $k$-th attempt chance:
$$
G_{s, k} =   \left \{  \begin{array}{lll}
 {\underline T}_s  +   \xi_s  & \mbox{ if  source $s^*=s$ transfers }  \\
 G_{s ,  k-1} + 1   &\mbox{ if none transfer either due to} \\   &\mbox{ collision or b/c none attempted }  \\
 G_{s,  k-1} + {\underline T}_{s^*}+1 & \mbox{if source $s^* \ne s$ transfers.}
 \end{array} \right .
$$
In the above, the transfer times $\{{\underline T}_s\}$ are special transfer times like in (\ref{Eqn_special_xi}), these are the transfer times given that the corresponding control SINR is the largest among all the agents, i.e., 
$$
{\underline T}_s = T_s \Big |\  \Big\{  \Phi^* = \Phi_s , \Phi_s > \theta \Big \}.
$$
This (plus $\xi_s$) is the age update at the end of successful  attempt  time slots, while age increases linearly in between  these time slots  (as in Figure \ref{fig_DNP}). 
Once again a renewal cycle can be formed (one for each source) with renewal epochs again being the  end of (${\underline T}$  length) time slots at which a message of the source is transferred.   The source gains access to transfer its message after possibly multiple collisions as according to CSMA protocol, as in previous section.   The renewal cycle is made of the time duration during which all these attempts were made, the other sources possibly transfer their messages (probably more than once also) and then the source transfers one message successfully.

\newcommand{\q}{{\bf q}}
\newcommand{\p}{{\bf p}}
\newcommand{\ps}{\phi}
\newcommand{\pss}{\phi}

\newcommand{\bps}{ {\bm \phi} }
\newcommand{\CN}{ {\mathcal N}   }
\newcommand{\cs}{cs}

As in the previous section, and with ${\mathcal N}_s$, ${\mathcal I}_{s, t}$ etc having the same meaning 
 and we again have the following for source s
(see (\ref{Eqn_Rc_multiple_source})):
\begin{eqnarray*}
{R_c}_{s,k} :=  R_{s,k} - R_{s, k-1}  & =&  \sum_{t=1}^{{\mathcal N}_s - 1} \left (  1+  \sum_{s' \ne s}    {\underline T}_{ s', t}    {\mathcal I}_{s', t}  \right ) +  1+  {\underline T}_{s, {\mathcal N}_s},    \\
\mbox{with } {\mathcal I}_{s, t} & =& \At_{s,t} 1_{\{ \Phi^*_t = \Phi_{s, t} , \Phi_{s, t} > \theta\} }
\end{eqnarray*}but with the following major differences: 
a) there  is  no waiting for message arrivals at source (i.e., no $\{\xi\}$), as this corresponds to saturated situation leading from one storage, but every attempt requires one unit time slot for control packet transfer;
b) there is a possibility of collision, because of which no source might transfer in all attempt slots, i.e., $\sum_s I_{s, t}$ need not be one always for any attempt slot $t$; and  more importantly
c)  the transfer times $\{{\underline T}_{s,  t} \}$ are not independent of source selection flags  $\{{\mathcal I}_{s , t}\}$.

To summarize  the renewal cycle of source $s$ has Geometric number  ${\mathcal N}_s$ of attempt slots,  
some of which are of length grater than one, due to message transfer by other sources.  Let 
$\{ {\mathcal M}_s^{s'} \}_{s' \ne s}$  represent  the number of these attempt slots  in which other sources transfer messages.  The renewal cycle can be represented as
(the attempt slot notations ($t$) are appropriately readjusted):
$$
{R_c}_{s,k}   =  {\mathcal N}_s + \sum_{s' \ne s}   \sum_{t=1}^{{\mathcal M}_s^{s'}}  {\underline T}_{s', t} +   {\underline T}_{s, {\mathcal N}_s}. 
$$
Define the following to represent the probability that the source $s$ succeeds in a typical attempt: 
\begin{eqnarray}
\pss_s  :=  P \Big  ( \left \{ \Phi_s =  \Phi^* \right \} \cap \left \{  \Phi_s > \theta \right \} \Big ) \mbox{ with } \Phi^* :=  \max_{s'} \Phi_{s'},
 \label{Eqn_success}
\end{eqnarray} 
and this  equals  the parameter  of Geometric ${\mathcal N}_s$. 
 The other sources can get access during this renewal period, i.e., during the continuous streak of  attempts in which source $s$ failed,  which happens with (conditional) probability\footnote{note  
 $$\Big ( \left \{  \Phi_{s'}  =   \Phi^*   \right \} \cap \left \{    \Phi^*  > \theta  \right \} \Big ) \cup   
 \Big ( \left  \{  \Phi_{s'}  =   \Phi^*   \right \} \cap \left \{    \Phi^*  > \theta  \right \}  \cap \left \{    \Phi_s  <  \theta  \right \}   \Big ) $$ 
 $$ = \left \{  \Phi_{s'}   =   \Phi^*   \right \} \cap \left \{    \Phi^*  > \theta  \right \} .$$ } for any $s' \ne s$
 $$\pss^{s'}_s  := P \Big (\left \{  \Phi_{s'}  =   \Phi^*   \right \} \cap \left \{    \Phi^*  > \theta  \right \} \Big |  
\left \{ \Phi_s <  \Phi^* \right \} \cup \left \{  \Phi_s < \theta \right \}  
  \Big ) = \frac{\pss_{s'}}{1- \pss_s}.$$  Further ${\mathcal M}_s^{s'} $ is the number of attempt slots successful for agent $s'$,  among ${\mathcal N}_s-1$  slots (before the renewal cycle of
  source $s$ ends) and thus
  $${\mathcal M}_s^{s'} = {\mathcal B} ({\mathcal N}_s-1,  \pss_s^{s'}),$$
 where  ${\mathcal B}$ is binomial random variable, depending upon ${\mathcal N}_s$. 
 As in previous sections AAoI is given by (\ref{Eqn_AAoI_by_RT_limit}) and we again need the first two moments  $E[{R_c}_s]$, $E[({R_c}_s )^2]$,  and $E[G_{s}]$.

As already mentioned, the transfer times $\{{\underline T}_s\}$ are  the special transfer times like in (\ref{Eqn_special_xi}), for example the first moment equals:
$$
E[{\underline T}_s] = E\left [ \frac{C}{\log (1+\Psi_s ) } \Big | \  \Phi_s = \Phi^* , \ \Phi_s > \theta   \right]  .
$$
By conditioning first on ${\mathcal N}_s$ then on $\{ {\mathcal M}_s^{s'} \}$ we have:
\begin{eqnarray}
E[{R_c}_s]  &=&  E[{\mathcal N}_s]  + \sum_{s' \ne s}  E[ {{\mathcal M}_s^{s'}} ]  E[  {\underline T}_{s'} ] +   E[ {\underline T}_{s} ] \nonumber \\
&=& \frac{1}{\pss_s}  + \sum_{s' \ne s}  \frac{1- \pss_s}{\pss_s} \pss^{s'}_{s}  E[  {\underline T}_{s'} ] +   E[ {\underline T}_{s} ] \nonumber \\
&=& \frac{1}{\pss_s}  + \sum_{s'}  \frac{\pss_{s'} }{\pss_s}    E[  {\underline T}_{s'} ] ,  \nonumber \\
E[G_s] &=&  E[{\underline T}_s] + \frac{1}{\lambda_s}.  \label{Eqn_RandCSMA_GsRc}
\end{eqnarray}
Further using  the corresponding conditional independence of quantities of  various attempt slots, we get (also note $I_{s'',t} I_{s',t} = 0$ when $s''\ne s'$):
\begin{eqnarray}
  E[({ R_c}_s )^2 ] &=&  E[ ({\mathcal N}_s)^2]   +  E[( {\underline T}_s)^2 ]   +  \sum_s \sum_{ s' \ne s}   E[ {{\mathcal M}_s^{s'}} ]  E[ ( {\underline T}_{s'} )^2 ]  \nonumber \\
 && \hspace{-19mm}  +   E[( {\mathcal N}_s  )^2 -    3{\mathcal N}_s  +2  ]  \left (   \sum_{ s' \ne s}  E[  {\underline T}_{s'} ] \right )^2   +  
2  \sum_{s'}  E[ ( {\mathcal N}_s -1)  {{\mathcal M}_s^{s'}}]   E[  {\underline T}_{s'} ]  \nonumber \\
&& \hspace{-19mm}  +   2 E[ {\mathcal N}_s   ] E[ {\underline T}_s ]  + 2 E[ {\mathcal N}_s  -1 ] E[ {\underline T}_s ]  \sum_{s' \ne s} E[ {\underline T}_{s'} ] .
\label{Eqn_RandCSMA_Rc2}
\end{eqnarray}
By  conditioning on ${\mathcal N}_s$ we get for each $s$:
\begin{eqnarray*}
E[\CN_s ] &=&  \frac{1}{\pss_s} \mbox{, }   \   \   \   E[\CN _s^2] =  \frac{2-\pss_s}{ \pss_s^2}  , \  \  \ 
%
E[{\mathcal M}_s^{s'}] =  E [{\mathcal N}_s - 1]  \pss_s^{s'}  =  \frac{ \pss_{s'}}{\pss_s} \\
&& \hspace{-12mm}
  \mbox{ and }
E[{\mathcal N}_s{\mathcal M}_s^{s'}]\ = \  E [{\mathcal N}_s ({\mathcal N}_s - 1)]   \pss_s^{s'} =  \frac{ 2\pss_{s'}}{\pss_s^2} .
\end{eqnarray*}
Then the  AAoI  of an agent  is given by  (\ref{Eqn_AAoI_by_RT_limit}) after substituting the above expressions.

 \subsubsection*{Rayleigh Channels}
We consider a special case of Rayleigh channels and compute $\{ \pss_s\}_s$. 
 When $\{H_{s,k}\}_k$ are  Rayleigh for each agent then $|H_s|^2$ is exponentially  distributed with parameter $1/(2\sigma_s^2)$, where $\sigma_s^2$ is the variance of the Gaussian terms, whose magnitude equals $|H_{s,k}|^2$.  
 
One can easily verify that

\vspace{-4mm}
{\small\begin{eqnarray*}
\left \{ \Phi_s  \ge  \Phi_{s'} \right \} &=&   \Bigg \{  H_s^4 \At_s  + H_s^2 \At_s  \left ( \sum_{s'' \ne s}    H_{s''}^2 \At_{s''}  +  \sigma_n^2 \right )  \\
&&  \hspace{5mm}
\ge   H_{s'}^4\At_{s'}  +H_{s'}^2 \At_{s'} \left (  \sum_{s'' \ne s'}    H_{s''}^2\At_{s''}  +   \sigma_n^2 \right )  \Bigg \}  \\
&=& \left \{   H_s^2 \At_s  \ge   H_{s'}^2 \At_{s'}  \right \}.
\end{eqnarray*}} 
 Thus the success parameter  equals:
\begin{eqnarray*}
\pss_s  =  P \left   ( \cap_{s \ne s'} \left \{   H_s^2 \At_s  \ge   H_{s'}^2 \At_{s'}  \right \}   \cap \left \{     H_s^2  \At_s   >   \theta  \sum_{s\ne s'}H_{s'}^2 \At_{s'}    + \theta \sigma_n^2   \right \}  \right ) .
\end{eqnarray*}
When $\theta \ge  1$, it is clear\footnote{By conditioning on  $H_j^2$ we get:
\begin{eqnarray}
P (H_s^2 >  \theta H_j^2 + c  )  = E[ e^{- (\theta H_{s'}^2 + c) / (2\sigma_s^2)} ]  =    e^{-0.5 c/ \sigma_s^2}  \frac{\sigma_s^2}{\sigma_s^2 + \theta  \sigma_{s'}^2}.
\end{eqnarray}}  that:
\begin{eqnarray*}
\pss_{s_1} & =&  P \left   (       H_{s_1}^2  \At_s   >   \theta  \sum_{s'\ne s_1}H_{s'}^2 \At_{s'}    + \theta \sigma_n^2  \right )   \\
&& \hspace{-12mm}=\  q_{s_1} E\Bigg [ e^{- \frac{ \theta  \sum_{s'\ne s_1}H_{s'}^2 \At_{s'}  + \theta \sigma_n^2 }{2 \sigma_{s_1}^2}}  \Bigg ], \mbox{    conditioning on $\sum_{s'\ne s_1}H_{s'}^2 \At_{s'} $}  \\
&  & \hspace{-12mm} = \ q_{s_1}  e^{-  0.5 \theta \sigma_n^2/ \sigma_{s_1}^2 }    \hspace{-5mm}
\sum_{{\bf j} \in \{0, 1\}^{S-1}}  \Pi_{l=2}^{S} (1-q_{s_l})^{1- j_l}  q_{s_l}^{j_l}  \left ( 1- j_l + j_l   \frac{ \sigma_{s_1}^2}{\sigma_{s_1}^2 + \theta \sigma_{s_{l}}^2} \right ), \\ 
&& \hspace{40mm}
\mbox{ by independence. }      
\end{eqnarray*}In the above product over empty sets (of sources) is defined to be 1.
For most of the practical scenarios, the systems operate at   nominal SINRs  for which $\sigma_n^2 $ is small.  Approximating $\sigma_n^2 \approx 0$ with $\theta < 1$, 
we have (when atleast one  $q_{s'} \ne 0$ with $s' \ne s$):
 \begin{eqnarray*}
\pss_s  &\approx &  P \left   ( \cap_{s \ne s'} \left \{   H_s^2 \At_s  \ge   H_{s'}^2 \At_{s'}  \right \}   \cap \left \{     H_s^2  \At_s   >   \theta  \sum_{s\ne s'}H_{s'}^2 \At_{s'}     \right \}  \right )  \\
&=& q_s P \left   ( \cap_{s \ne s'} \left \{   H_s^2  \ge   H_{s'}^2 \At_{s'}  \right \}   \right )  \mbox{, conditioning on } H_s^2 \\
&=& q_s   E \left [  \Pi_{s' \ne s} \left (   1-q_{s'}  + q_{s'}  \left (1 -  e^{- 0.5 H_s^2 / \sigma_{s'}^2}  \right ) \right ) \right ]\\
&=& q_s   E \left [  \Pi_{s' \ne s} \left (   1-  q_{s'}    e^{- 0.5 H_s^2 / \sigma_{s'}^2}  \right )   \right ] \mbox{ and } \\
\pss_s &=&    q_s e^{- 0.5 \theta \sigma_n^2 /\sigma_s^2 } \mbox{ when }  q_{s'} = 0 \mbox{ for all }  s' \ne s.
 \end{eqnarray*}
In the same way, one can obtain the conditional moments $E[\underline{T}_s]$ and $E[ (\underline{T}_s )^2]$ based on the range of $\theta$.
For example, 
\begin{eqnarray*}
E[{\underline T}_{s_1} ] &=& \frac{1}{\pss_{s_1} }  E \left  [  T_s ;   H_{s_1}^2  \At_s   >   \theta  \sum_{s'\ne s_1} H_{s'}^2 \At_{s'}    + \theta \sigma_n^2  \right  ]    \\
&& \hspace{-20mm} = \ \frac{q_{s_1}}{\pss_{s_1} }  \sum_{{\bf j} \in \{0, 1\}^{S-1}}  \Pi_{l=2}^{S} (1-q_{s_l})^{1- j_l}  q_{s_l}^{j_l}   
 E \left  [  T_s ;   H_{s_1}^2    >   \theta  \sum_{s_l \ne s_1} H_{s_l}^2 j_{l}    + \theta \sigma_n^2  \right  ] .
\end{eqnarray*}
 The expected values in all these expressions do not depend upon ${\bf q}$, the attempt probabilities, the parameters to be tuned for optimal design. Nevertheless, one needs  to know/ estimate the terms 
 $$
\left \{  E \left  [  T_s ;   H_{s_1}^2    >   \theta  \sum_{s_l \ne s_1} H_{s_l}^2 j_{l}    + \theta \sigma_n^2  \right  ] \right   \}_{{\bf j} \in \{0, 1\}^{S-1}}
 $$  to solve the relevant optimization / game theoretic problems.

\subsubsection*{\underline{Two sources}} For the case with two sources, i.e., with $S=2$ the above expressions simplify as below.
When $\theta \ge 1$,
\begin{eqnarray*}
E[{\underline T}_{s_1} ] &=&\frac{q_{s_1}}{\pss_{s_1} }    \Big (   (1- q_{s_2} )
E \left  [  T_s ;   H_{s_1}^2    >    \theta \sigma_n^2  \right  ]  \\
&& +   q_{s_2}  E \left  [  T_s ;   H_{s_1}^2    >   \theta  H_{s_2}^2     + \theta \sigma_n^2  \right  ]
\Big  )   .
\end{eqnarray*}
When $\theta < 1$  and using $\theta \sigma_n^2  \approx 0$  only for terms with $q_{s_2} \ne 0$ we get
\begin{eqnarray*}
E[{\underline T}_{s_1} ] \approx  \frac{q_{s_1}}{\pss_{s_1} }    \left (   (1- q_{s_2} )
E \left  [  T_s  ;    H_{s_1}^2    >    \theta \sigma_n^2 \right  ]  +   q_{s_2}  E \left  [  T_s ;   H_{s_1}^2    >      H_{s_2}^2       \right  ]
\right ) .
\end{eqnarray*}
Thus by considering moderate SINRs, one can approximate for any $\theta$ as below
\begin{eqnarray*}
E[{\underline T}_{s_1} ] &\approx &\frac{q_{s_1}}{\pss_{s_1} }    \Big (   (1- q_{s_2} )
E \left  [  T_s ;    H_{s_1}^2    >    \theta \sigma_n^2  \right  ]   \\ 
&& \hspace{10mm} +   q_{s_2}  E \left  [  T_s ;   H_{s_1}^2    >  \max \{1, \theta\}    H_{s_2}^2       \right  ]
\Big )  \\
E[({\underline T}_{s_1} )^2 ] &\approx &\frac{q_{s_1}}{\pss_{s_1} }    \Big  (   (1- q_{s_2} )
E \left  [ ( T_s )^2 ;    H_{s_1}^2    >    \theta \sigma_n^2 \right  ]  \\
&& \hspace{10mm}  +   q_{s_2}  E \left  [  (T_s )^2 ;   H_{s_1}^2    >  \max \{1, \theta\}    H_{s_2}^2       \right  ]
\Big )  \\
\pss_{s_1}  &\approx & q_{s_1} \left ( 1-q_{s_2} \frac{\max \{\theta, 1\} \sigma_{s_2}^2 }{\max \{\theta, 1\} \sigma_{s_2}^2  + \sigma_{s_1}^2  }  \right ).
\end{eqnarray*}
Substituting these expressions in (\ref{Eqn_AAoI_by_RT_limit}) one can compute the AAoIs ${\bar a}_1 ({\bf q}), {\bar a}_2 ({\bf q})$ for any vector of attempt probabilities 
${\bf q}= (q_1, q_2)$.

\subsection*{Non-cooperative Game}
One of the important issues in the design of CSMA protocol is the design of attempt probabilities ${\bf q} := (q_1 \cdots, q_S)$.  This results in a natural game theoretic setting, each agent tries to improve its own performance (e.g., AAoI). 
AAoI is a relatively new performance metric, while previously the performance metric,
throughput, is widely used. Throughput can be defined as {\it the fraction of times a source gains access to transfer its packet} and it is easy to verify that the   throughput of (saturated) source $s$ equals 
$$q_s \Pi_{s' \ne s} (1-q_{s'}). $$  
 It is well known  that for the non-cooperative game corresponding to CSMA protocol   with throughput 
as utility and without considering the cost of attempts, the Nash Equilibrium (NE)\footnote{A Nash equilibrium is a  profile of strategies  where no player
can improve its reward by deviating unilaterally from the  strategy specified in the NE.}  is all ones, i.e., ${\bf q}^* = (1, \cdots, 1)$. 
However it is interesting to observe that the throughput of any source at this  NE equals 0.   
Thus designers consider cooperative solutions.  We will notice that we have a drastically different result when Freshness of information is considered.

\subsection*{Numerical Observations}  
Let ${\bf q} =  (q_1, q_2, \cdots,  q_S)$  be the vector of attempt probabilities and let ${\bar a}_s ({\bf q})$ the AAoI of agent $s$. The AAoI  ${\bar a}_s ({\bf q})$ equals
(\ref{Eqn_AAoI_by_RT_limit}) after substituting   (\ref{Eqn_RandCSMA_GsRc})-(\ref{Eqn_RandCSMA_Rc2}) and the corresponding terms. The closed form expressions are derived for all the terms describing 
(\ref{Eqn_RandCSMA_GsRc})-(\ref{Eqn_RandCSMA_Rc2}), but for the conditional moments, 
$E[{\underline T}_s ]$,  $E[({\underline T}_s)^2 ]$ etc.
 One needs to estimate  these moments   numerically   and then  AAoI can be computed.

We  consider the following strategic form non cooperative game:
$$ \left  < \{1, \cdots, S\},  \{{\bf q} \in [0,1]^S\},   \{{\bar a}_s , s\in S\} \right >  $$ with  AAoI as performance.  
We also consider cooperative solution when all the agents jointly minimize a social choice function, the average of the AAoI across all agents:
\begin{eqnarray*}
\min_{ {\bf q} }  \frac{1}{S} \sum_s {\bar a}_s ({\bf q})  . 
\end{eqnarray*}

\begin{table}[h!]
\centering
\caption{With  $S=2$  agents, with ${\bar a} := ({\bar a}_1 + {\bar a}_2 )/2$}
\label{M=2 }
\begin{tabular}{|c|c|c|c|c|c|c|}
\hline
\multirow{2}{*}{$\theta$} & \multicolumn{3}{c|}{Nash equilibrium} & \multicolumn{3}{c|}{Cooperative  Solution} \\ \cline{2-7} 
                  &  $q_1^*$     &$q_2^*$       & ${\bar a}^*$      & $q^*_1$      & $q^*_2$      &  ${\bar a}^*$     \\ \hline
                 0.1 & 1      & 1      & 5.61      & 1      & 1       & 5.61      \\ \hline
                0.5  & 1      & 1      & 5.58      & 1       & 0.92       & 5.54       \\ \hline
                0.6  & 1      & 1      & 5.55      & 1       & 0.92       & 5.51       \\ \hline
                
                0.8  & 1      & 1      & 5.53      & 1       & 0.92       & 5.47       \\ \hline
                0.9  & 1      & 1      & 5.52      &1       & 0.92      & 5.46      \\ \hline
                 1.2 & 1      & 1      & 5.76      & 1     & 0.82      & 5.60      \\ \hline
\end{tabular}
\end{table}
We begin with a two agent example in in Table \ref{M=2 }, which   shows the equilibrium solution  for different values of the parameter $\theta$. We  set  $ S = 2 $,   $\sigma^2=(6,10)$,   $C=(5,5)$ and noise variance $\sigma^2_n =0.5$.  
We have tabulated the estimates of NE, cooperative solution and the average AAoI at respective equilibria. 
 Interestingly the NE ($(q_1^*, q_2^*)$ of   second and third columns) as well as the cooperative solution  ($(q_1^*, q_2^*)$ of fourth and fifth columns) are both  close to $(1,1)$.  More interestingly the {\it performance at 
 NE $(1,1)$ is almost equal to that with cooperative solution, when $\theta $ is small}. 

A second example with  $S=6$ agents is considered in Table \ref{M=6}. We consider a symmetric scenario where for each source  $\sigma^2_s=10$,  $C=1$ and noise variance $\sigma^2_n =2$.  Because of symmetry we tabulated only one component of NE as well as the cooperative solution. 
We observe again that $(1, \cdots, 1)$ is NE as well as the cooperative solution for small values of threshold $\theta $ (the first two rows with $\theta \le 1$). While the cooperative solution is largely different from NE and the average AAoI is improved with cooperative solution for large $\theta$.  However and more importantly, the average AAoI is the best (7.62) with small $\theta$. Thus with large number of agents it is optimal to consider  
 low   threshold (the minimum value at which the signal can be detected) and each source should attempt  with probability one.  With this, the NE as well as cooperative solution is the same, further they provide the best average AAoI.  

 Further since $(1, \cdots, 1)$ is cooperative solution as well as the NE, the players have no incentive to play differently.  Thus the conclusions (cooperative and NE are almost similar) with AAoI as performance measure is drastically different in comparison with the case when throughput is considered.  If one also considers SINR dependent throughput,  and
 if the performance measure  is the time average of the  throughput, we may  have similar conclusions with Random CSMA. This could be an interesting direction to explore in future.

 This precisely is like the  `reverse bus paradox' that we mentioned in the beginning: with more independent choices (number of agents is high plus  with $\theta$ small, more agents are heard by the destination),  and with   the algorithm picking the best,  update packets are transferred for shorter duration with high probability. Thus the AAoI of all the agents improves simultaneously in most of the scenarios.

\begin{table}[h!]
\centering
\caption{With S=6 agents}
\label{M=6}
\begin{tabular}{|c|c|c|c|c|}
\hline
\multirow{2}{*}{$\theta$} & \multicolumn{2}{c|}{Nash Equilibrium } & \multicolumn{2}{l|}{Cooperative Solution} \\ \cline{2-5} 
                 & $q$   &  ${\bar a}^*$        &     $q^*$     & ${\bar a}^*$    \\ \hline
                 0.01  &    1       & 7.62          & 1          & 7.62          \\ \hline
                 0.1  &    1       & 7.65           &1           &  7.63          \\ \hline
                 0.34  &    1       & 7.84         &0.82         & 7.80          \\
 \hline
               0.4  &    1       & 8.38           &0.73           & 8.01          \\ \hline
                 0.5           &  1         &10.04           &0.55           &   8.39        \\ \hline
                1  &    1       & 36.67           &0.37           & 10.52          \\ \hline

\end{tabular}

\end{table}

\section{Conclusions}

We considered the problem  of keeping  the destination regularly updated with  fresh  information. 
Old information can   become completely obsolete once a new update is received, when the  focus is entirely about fresh information.  
The systems naturally become lossy, in the sense, some packets would  be discarded.  
We developed a methodology to study the freshness of information, using average age of information (AAoI) as performance metric, for lossy systems. A packet at destination can automatically be discarded once a new update is available. However a new packet at source, while the source is in the middle of transfer of a older packet, demands an important decision: which packet to be discarded. Older packets  can be transferred faster to the destination, while the new packet may have fresh information but would require more time to reach the destination.  It may be wiser to base these decisions  on the state of the system,  the age of the previous update of the same information at destination,  at the decision epoch.   However  two static policies, drop always the new packets (DNP) or drop always the old packets (DOP),  are optimal/almost optimal among a certain class of  stationary Markov policies. 
Further   DOP  is the best for small update rates, and DNP is best for large update rates for any given distribution of the transfer times.  
If the variance of the transfer times is higher,  DOP becomes optimal  for a wider range of update rates.

When multiple sources are to update their respective information at the same destination using the same medium, there exists an inherent competition. 
The sources attempt to regularly update their information over common wireless channel, using  CSMA (carrier-sense multiple access) protocol.    
We derived the performance, AAoI for each source depending upon the actions (attempt probabilities) of different sources.   The natural dropping of `bad' packets (only the ones with good SINR are selected to transfer their message in each attempt  slot)   in CSMA protocol   ensures  that the non-cooperative solution also   optimizes/ almost optimizes a  social choice function. 
We observe that the profile of attempt probabilities that form the Nash equilibrium for Random CSMA environment  also minimize the sum of the AAoI of all the  agents.

 \bibliographystyle{plain}

\section*{Appendix A: Proofs}

{\bf Proof of Theorem \ref{Thm_dn_lt_do}:}
As a first step, one can easily observe
that the coefficients of the lower bound function $f_o$ depend upon $\theta$ only via the stationary distribution $\pi_\theta$, in particular only via $\pi_\theta(0)$, 
i.e., $f_o (\theta) = f_o (\pi_\theta(0) )$.
Further
 the function $\theta \mapsto \pi_\theta(0)$ is ONTO (see (\ref{Eqn_pi})) and  hence one can equivalently optimize $f_o$ using $\pi :=\pi_\theta(0)$: 
$$
f_o (\theta) = f_o (\pi) =  \frac{  d_o  \bigg ( (b_n - b_o )\pi  +  b_o  \bigg ) + 0.5( c_n -c_o)  \pi +  0.5 c_o  }{(d_n -d_o)\pi + d_o }. 
$$
 The first derivative for the lower bound function is:
\begin{eqnarray}
   f_o^\prime(\pi)  
   &=& \frac{ 0.5 (c_n d_o - c_od_n) +d_o ( b_n   d_o -   b_o d_n )   }{ (\pi(d_n-d_o)+d_o ) ^2} . \label{Eqn_first_derivative_ERc} 
\end{eqnarray}
From (\ref{Eqn_cndo_minus_dnco}) of Appendix A:
\begin{eqnarray*}
c_n d_o - c_o d_n  
&  =&  
\frac{ E[T^2]}{ \lambda \gamma }    
 +  \left (\frac{1}{\lambda} +  E[T]  \right )  \left (  E[T e^{-\lambda T}]  - \frac{ (1-  \gamma) }{\lambda  } \right )  
 \frac{2}{\lambda \gamma^2}.
\end{eqnarray*}
Thus the numerator of the derivative (\ref{Eqn_first_derivative_ERc}) is proportional to,
\begin{eqnarray*}
c_n d_o - c_o d_n  + 2 d_o (b_n d_o - b_o d_n)  \hspace{-32mm} \\
&  =&  
\frac{ E[T^2]}{ \lambda \gamma }    
 +  \left (\frac{1}{\lambda} +  E[T]  \right )    E[T e^{-\lambda T}]  \left (\frac{2}{\lambda \gamma^2} -  2 \frac{1}{\lambda \gamma^2 }\right )  \\
 &&   -    \frac{ 1 }{\lambda \gamma  }  \left (  \left (\frac{1}{\lambda} +  E[T]  \right )    \frac{2(1-\gamma)}{\lambda \gamma} - E[T] \frac{2}{\lambda \gamma} \right )   \\
  &=&  \frac{ E[T^2]}{ \lambda \gamma }      + \frac{2}{\lambda^2 \gamma } (E[T] - E[\Rcc]  ) > 0,  \mbox{ when $d_n \ge d_o$. }
\end{eqnarray*}
Thus the derivative  $f_o' (\theta) > 0$ for all $\theta$, hence the lower bound $f_o$ is increasing with $\pi$, and thus the unique minimizer of $f_o$ is at  $\pi^* = 0$. This   implies the DOP scheme (see (\ref{Eqn_pi})) is optimal for AAoI ${\bar a}(\centerdot)$.   \eop

 {\bf Proof of Theorem \ref{Thm_dn_gt_do}:}  As before it suffices to show that the numerator of derivative of $f_n$  (with respect to  $\pi$) is negative. 
Recall the following:
\begin{eqnarray*}
c_n d_o - c_o d_n  
&  =&  
\frac{ E[T^2]}{ \lambda \gamma }    
 +  \left (\frac{1}{\lambda} +  E[T]  \right )  \left (  E[T e^{-\lambda T}]  - \frac{ (1-  \gamma) }{\lambda  } \right )  
 \frac{2}{\lambda \gamma^2} , \\
 d_o &=&  \frac{1}{\lambda \gamma }  , \  \   b_o = \frac{E[ T e^{-\lambda T}]} {\gamma},   \  \  d_n = \frac{1}{\lambda }  + E[T]
\end{eqnarray*}
The numerator of derivative of $f_n$  is proportional to,
\begin{eqnarray*}
c_n d_o - c_o d_n  + 2 d_n (b_n d_o - b_o d_n)  \hspace{-38mm} \\
&  =&  
\frac{ E[T^2]}{ \lambda \gamma }    
 +  \left (\frac{1}{\lambda} +  E[T]  \right )    E[T e^{-\lambda T}]  \left (\frac{2}{\lambda \gamma^2} - \frac{ 2}{\gamma} \left ( \frac{1}{\lambda}  + E[T] \right ) \right )  \\
 &&   -    \frac{ 2 }{\lambda \gamma  } \left (\frac{1}{\lambda} +  E[T]  \right )    \left (     \frac{ 1-\gamma}{\lambda \gamma} - E[T]   \right )   \\
 &  =&  
\frac{ E[T^2]}{ \lambda \gamma }    
 - \frac{ 2}{\lambda \gamma } \left (\frac{1}{\lambda} +  E[T]  \right )   \left ( 1-  \lambda  E[T e^{-\lambda T}] \right )    (d_o - d_n ).
\end{eqnarray*} Thus the theorem.  \eop

\section*{ Some useful terms}

The estimate of  the term $c_n d_o - d_n c_o$:
{\small \begin{eqnarray}
c_n d_o - d_n c_o &=&  \left (\frac{2}{\lambda^2 } + E[T^2] + \frac{2 E[T] }{\lambda} \right ) \left ( \frac{1}{\lambda} + \frac{1-\gamma}{\lambda \gamma} \right )
\nonumber \\
&& \hspace{-24mm}-  \left (\frac{1}{\lambda} +  E[T]  \right ) \left ( \frac{2}{\lambda^2 }  +  \frac{ 2 (1-  \gamma)  }{\lambda^2 \gamma} 
- \frac{2 E[T e^{-\lambda T}]}{ \lambda\gamma^2}   + 2 \frac{  (1- \gamma)^2  }{\lambda^2 \gamma^2 } +\frac{2 (1-  \gamma) }{\lambda^2 \gamma}  \right )
\nonumber  \\ \nonumber  \\
&& \hspace{-24mm}  =\left (\frac{2}{\lambda^2 } + E[T^2] + \frac{2 E[T] }{\lambda} \right ) \left ( \frac{1}{\lambda \gamma}   \right ) \nonumber 
\\&& \hspace{-10mm}-  \left (\frac{1}{\lambda} +  E[T]  \right ) \left ( \frac{2}{\lambda^2 \gamma  }     \nonumber 
- \frac{2 E[T e^{-\lambda T}]}{ \lambda\gamma^2}   + 2 \frac{  (1- \gamma)^2  }{\lambda^2 \gamma^2 } +\frac{2 (1-  \gamma) }{\lambda^2 \gamma}  \right ) \\ \nonumber  \\
&& \hspace{-24mm} =  E[T^2]\left ( \frac{1}{\lambda \gamma}   \right )  \nonumber 
 -  \left (\frac{1}{\lambda} +  E[T]  \right ) \left (    
- \frac{2 E[T e^{-\lambda T}]}{ \lambda\gamma^2}   + 2 \frac{  (1- \gamma)^2  }{\lambda^2 \gamma^2 } +\frac{2 (1-  \gamma) }{\lambda^2 \gamma}  \right ) \\
\nonumber   \\
&& \hspace{-24mm}  =  \frac{1}{\gamma} \left (  
\frac{ E[T^2]}{ \lambda}    
 +  \left (\frac{1}{\lambda} +  E[T]  \right )  \left ( \frac{2 E[T e^{-\lambda T}]}{ \lambda \gamma }  \right )  \right )  \nonumber  \\
&&   -   \frac{1}{\gamma} \left (   \left (\frac{1}{\lambda} +  E[T]  \right )   \left (  \frac{2 (1-  \gamma) }{\lambda^2 \gamma } \left ( (1- \gamma)+  \gamma \right )  \right )  \right ) \nonumber  \\
&& \hspace{-24mm}  =  \frac{1}{\gamma} \left (  
\frac{ E[T^2]}{ \lambda}    
 +  \left (\frac{1}{\lambda} +  E[T]  \right )  \left ( \frac{2 E[T e^{-\lambda T}]}{ \lambda \gamma  }  \right )     
   -     \left (  \frac{1}{\lambda} +  E[T]  \right )     \frac{2 (1-  \gamma) }{\lambda^2 \gamma }   \right )  \nonumber  \\
  \nonumber  \\
&& \hspace{-24mm}  =  
\frac{ E[T^2]}{ \lambda \gamma }    
 +  \left (\frac{1}{\lambda} +  E[T]  \right )  \left (  E[T e^{-\lambda T}]  - \frac{ (1-  \gamma) }{\lambda  } \right )  
 \frac{2}{\lambda \gamma^2} .   \label{Eqn_cndo_minus_dnco}
 \end{eqnarray}}

By appropriate conditioning:
\begin{eqnarray}
E \Big [\xi_s ;  \xi_s \le \min_{s' \ne s}  \xi_s'  \Big  ]   & = & E\Big [  \xi_s e^{- (\lamS- \lambda_s)  \xi_s}  \Big ]  =  \frac{ \lambda_s}{\lamS^2} \\
E \Big [\xi_s^2 ;  \xi_s \le \min_{s' \ne s}  \xi_s'  \Big  ]  & =  & E\Big [  \xi_s^2 e^{- (\lamS- \lambda_s)  \xi_s}  \Big ]  =  \frac{2 \lambda_s}{\lamS^3}
\end{eqnarray}

\Detail{Further by conditioning on $T$:
\begin{eqnarray*}
E[\xi; \xi \le T ] = E[  e^{-\lambda T} (T + 1/ \lambda )  - 1/  \lambda ] =  E[T e^{-\lambda T}]  - \frac{ (1-  \gamma) }{\lambda  },
\end{eqnarray*}
and 
 thus we have:
 \begin{eqnarray}
\label{Eqn_cndo_minus_dnco1}
c_n d_o - d_n c_o &=&
\frac{ E[T^2]}{ \lambda \gamma }    
 -  \left (\frac{1}{\lambda} +  E[T]  \right )   E[\xi ; T \ge \xi] 
 \frac{2}{\lambda \gamma^2}   
      . \hspace{5mm}
\end{eqnarray}}{}

\medskip
\medskip

\end{document}